%% file: main.tex
\documentclass[]{aiaa-tc}

\usepackage{amsmath}
\usepackage{amssymb}
\usepackage{verbatim}
\usepackage{subfigure}
\usepackage{placeins}
\usepackage{overpic}
\usepackage{multicol}

\usepackage[pdftex]{hyperref}
\hypersetup{
pdfauthor   = {},
pdftitle    = {},
pdfsubject  = {},
pdfkeywords = {},
pdfcreator  = {PDFLaTeX with hyperref package},
pdfproducer = {pdflatex},
colorlinks  = true,
linkcolor   = blue,
anchorcolor = red,
citecolor   = blue,
filecolor   = red,
urlcolor    = red
}

\title{Large Eddy Simulations of Supersonic Jet Flows for Aeroacoustic Applications}

\author{
Carlos Junqueira-Junior\thanks{Ph.D. Student,
Graduate Program on Computer Sciences and Electrical Engineering,
Departamento de Ci\^{e}ncia e Tecnologia Aeroespacial, DCTA/ITA;
E-mail: junior.hmg@gmail.com.}\\
{\normalsize\itshape Instituto Tecnol\'{o}gico de Aeron\'{a}utica, 12228-900
S\~{a}o Jos\'{e} dos Campos, SP, Brazil}\\
\and
Sami Yamouni\thanks{Postdoctoral Reasearch Fellow,
Graduate Program on Computer Sciences and Electrical Engineering,
Departamento de Ci\^{e}ncia e Tecnologia Aeroespacial, DCTA/ITA;
E-mail: sami.yamouni@gmail.com.}\\
{\normalsize\itshape Instituto Tecnol\'{o}gico de Aeron\'{a}utica, 12228-900
S\~{a}o Jos\'{e} dos Campos, SP, Brazil}\\
\and
Jo\~{a}o Luiz F. Azevedo
\thanks{Senior Research Engineer, Aerodynamics Division,
Departamento de Ci\^{e}ncia e Tecnologia Aeroespacial, DCTA/IAE/ALA;
E-mail: joaoluiz.azevedo@gmail.com. AIAA Fellow.}\\
{\normalsize\itshape Instituto de Aeron\'{a}utica e Espa\c{c}o, 12228-903
S\~{a}o Jos\'{e} dos Campos, SP, Brazil}
\and
William R. Wolf
\thanks{Assistant Professor, Faculty of Mechanical Engineering;
E-mail: wolf@fem.unicamp.br. AIAA Member}\\
{\normalsize\itshape Universidade Estadual de Campinas, 13083-970
Campinas, SP, Brazil}
}

 \AIAApapernumber{2015-NUMBER}
 \AIAAconference{22nd AIAA Computational Fluid Dynamic Conference, June, 2015, Dallas, Texas}
 \AIAAcopyright{\AIAAcopyrightD{2015}}

\begin{document}

\maketitle


\begin{abstract}

\section*{Abstract}

Current design constraints have encouraged the studies of 
aeroacoustics fields around compressible jet flows. The 
present work addresses the numerical study of unsteady 
turbulent jet flows for aeroacoustic analyses 
of main engine rocket plumes. A novel large eddy simulation 
(LES) tool is developed in order to reproduce high fidelity 
results of compressible jet flows which could be used for aeroacoustic 
studies with the Ffowcs Williams and Hawkings approach. The 
numerical solver is an upgrade of an existing Reynolds-averaged 
Navier-Stokes solver previously developed in the group. The 
original framework is rewritten in a modern fashion and 
intensive parallel computation capabilities have been added 
to the code. The LES formulation is written using the finite 
difference approach. The energy equation is carefully discretized 
in order to model the energy equation of the filtered Navier-Stokes 
formulation. The classical Smagorinsky model is the chosen 
subgrid scale closure for the present work. Numerical simulations 
of perfectly expanded jets are performed and compared with the 
literature in order to validate the new solver. Moreover, speedup 
and the computational performance of the code are evaluated and 
discussed. 
Flow results are used for an initial evaluation of 
the noise radiated from the rocket plume.

\end{abstract}


\input{sources/nomenclature/nomenclature}


\section{Introduction}

One of the main design issues related to launch vehicles 
lies on noise emission originated from the complex 
interaction between the high-temperature/high-velocity 
exhaustion gases and the atmospheric air. These emissions 
yield very high noise levels, which must be minimized due
to several design constraints. For instance, the 
resulting pressure fluctuations can damage the solid 
structure of different parts of the launcher by 
vibrational acoustic stress. Therefore, it is a design
constraint to consider the loads resulting from acoustic 
sources in the structural dimensioning of large launch 
vehicles during the take 
off and also during the transonic flight. Moreover,
one cannot neglect the energy dissipation effect 
caused by the acoustic waves generated even if the 
vehicles is far from the ground. Theoretically, all 
chemical energy should be converted into kinectic
energy. However, in reallity, the noise generation 
consumes part of the chemical energy.

The acoustic design constraints have encouraged the 
studies of aeroacoustic fields around compressible jet 
flows. Instituto de Aeronautica e Espa\c{c}o (IAE) 
in Brazil is interested in this flow configuration for 
rocket design applications. Unsteady property fields of 
the flow are necessary for the aerocoustic studies. 
Therefore, the present work addresses the development
of a new numerical tool for the study of unsteady turbulent 
compressible jet flows for aeroacoustic applications. 
The novel code uses the large eddy simulation (LES) formulation 
due to recent successful results of such approach for similar flow 
configurations \cite{Bodony05i8, Garnier09}. Such formulation is 
able to reproduce high fidelity results for compressible jet flows, 
which, in turn, could be used for aeroacoustic studies using 
the Ffowcs Williams and Hawkings approach \cite{Wolf2012}.

The solver is an upgrade of a finite difference Reynolds-averaged 
Navier-Stokes (RANS) solver developed to study turbulent flows 
for aerospace applications \cite{BIGA02}. Generally, LES calculations
demand very dense grids. Hence, 
high performance computing (HPC) is a requirement for such 
simulations. Dynamic memory allocation and parallel 
computation capabilities have been added to the legacy code 
\cite{BIGA02}. The communication between processors 
is performed by message passing interface (MPI) protocols. The 
CFD general notation system (CGNS) is also included in the numerical 
tool. The LES formulation is written using the finite difference 
approach. Inviscid numerical fluxes are calculated using a 
second-order accurate centered scheme with the explicit addition 
of artificial dissipation. A five-step second-order accurate 
Runge-Kutta scheme is the chosen time marching method. 

The System I formulation \cite{Vreman1995} is used here in 
order to model the filtered terms of the energy equation. 
The Smagorinsky \cite{Smagorinsky63} model is the chosen  model 
to calculate the components of the subgrid scale (SGS) tensor. 
The Eidson \cite{eidson85} hypothesis is the chosen model to 
calculate the subgrid scale terms of the filtered energy equation. 
For the sake of simplicity, only perfectly expanded jet configurations 
are studied in the current work. The LES tool is validated through 
a comparison of results of perfectly expanded jets with numerical 
\cite{Bodony05i8, Mendez10} and experimental 
\cite{Tanna77, bridges2008turbulence} data. Moreover, the 
computational performance of the solver is evaluated and discussed.
Flow results are used for an initial evaluation of 
the noise radiated from the rocket plume.


  \input{sources/theoretical/theoretical}

\FloatBarrier


  \input{sources/turb_mod/turb_mod}

\FloatBarrier


  \input{sources/coord/coord}

\FloatBarrier


  \input{sources/dimlss/dimlss}

\FloatBarrier


  \input{sources/numerical/numerical}



  \input{sources/hpc/hpc}

\FloatBarrier


  \input{sources/results/results}

\FloatBarrier


  \input{sources/conclusion/conclusion}

\section*{Acknowledgments}

The authors gratefully acknowledge the partial support for this research provided by
Conselho Nacional de Desenvolvimento Cient\'ifico e Tecnol\'ogico, CNPq, under the
Research Grants No.\ 309985/2013-7\@. This work is also supported 
by Funda\c{c}\~{a}o Coordena\c{c}\~{a}o de Aperfei\c{c}oamento de Pessoal de N\'{\i}vel 
Superior, CAPES, through a Ph.D. scholarship for the first author. The authors are also 
indebted to the partial financial support received from Funda\c{c}\~{a}o de Amparo 
\`{a} Pesquisa do Estado de S\~{a}o Paulo, FAPESP, under the Research Grants 
No.\ 2013/07375-0 and No.\ 2013/21535-0. This work has been conducted with
the computational resources of the National Supercomputing Center (CESUP), from 
Universidade Federal do Rio Grande do Sul, UFRGS, in Brazil.


\bibliography{sources/references}
\bibliographystyle{aiaa}

\end{document}

%% file: sources/nomenclature/nomenclature.tex

\section*{Nomenclature}

\begin{multicols}{2}

\subsection*{Abbreviations}

1-D: One dimensional\\
2-D: Two dimensional\\
3-D: Three dimensional\\
CFD: Computational fluid dynamics\\
CFL: Courant-Friedrichs-Lewys number \\
CGNS: CFD general notation system\\
EDQNM: Eddy-damped quasi-normal Markovian\\
IAE: Instituto de Aeron\'{a}utica e Espa\c{c}o\\
LES: Large eddy simulation\\
MPI: Message passing interface\\
SGS: Subgrid scale

\subsection*{English Characters}

$a$: Speed of sound\\
$B_{i}$: Subgrid terms of energy equation \\
$C_{p}$: Specific heat at constant pressure\\
$C_{s}$: Smagorinsky constant\\
$C_{v}$: Specific heat at constant volume\\
$d$: Inlet diameter\\
$\mathbf{d}$: Artificial dissipation term\\
$\mathcal{D}$: SGS viscous diffusion term\\
$E$: Total energy per mass unity\\
$\mathbf{E}_{e}$: Inviscid flux vector in the axial direction\\
$\mathbf{E}_{v}$: Viscous flux vector in the axial direction\\
$\mathbf{F}_{e}$: Inviscid flux vector in the radial direction\\
$\mathbf{F}_{v}$: Viscous flux vector in the radial direction\\
$\mathbf{G}_{e}$: Inviscid flux vector in the azimuthal direction\\
$\mathbf{G}_{v}$: Viscous flux vector in the azimuthal direction\\
$J$: Jacobian of the coordinate transformation\\
$l$: Reference lenght\\
$M$: Mach number\\
$p$: Static pressure\\
$Pr$: Prandtl number\\
$q$: Heat flux vector\\
$\mathcal{Q}_{j}$: Components of SGS temperature flux\\
$R$: Gas constant\\
$\mathbf{R}$: Riemman invariant\\
$Re$: Reynolds number\\
$RHS$: Right hand side of the equation\\
$S_{1}$: Sutherland constant\\
$S_{ij}$: Components of rate-of-strain tensor\\
$Sp_{N}$: Speedup for N processors\\
$t$: time\\
$T$: Temperature\\
$T_{N}$: CPU time using N processors\\
$T_{s}$: CPU time using single processor\\
$\textbf{u}$: Velocity vector\\
$u$: Components of the velocity vector in a Catesian coordinate system\\
$U$: Component of the contravariant velocity in the axial direction\\
$V$: Component of the contravariant velocity in the radial direction\\
$\vec{n}$ : Unity outward normal vector\\
$W$: Component of the contravariant velocity in the azimuthal direction\\
$\mathcal{W}$: Modified conservative properties of Turkel and Vatsa artificial dissipation\\
$\textbf{x}$: Spacial coordinates of a Cartesian coordinate system\\

\subsection*{Greek Characters}

$\alpha_{l}$: Runge-Kutta constants\\
$\beta_{j}$: Viscous terms of energy equation\\
$\gamma$: Specific heat ratio\\
$\Delta t$: Time-step\\
$\mathcal{E}$: SGS viscous dissipation \\
$\epsilon$: Turkel and Vatsa artificial dissipation terms \\
$\zeta$: Azimuthal direction\\
$\eta$: Radial direction\\
$\kappa$: Thermal conductivity\\
$\lambda$: Spectral radius-based scaling factor\\
$\lambda_{\zeta}$: Spectral radii in the $\zeta$ direction\\
$\lambda_{\eta}$: Spectral radii in the $\eta$ direction\\
$\lambda_{\xi}$: Spectral radii in the $\xi$ direction\\
$\mu$: Dynamic viscosity\\
$\nu$: Kinematicc viscosity\\
$\xi$: Axial direction\\
$\Pi_{dil}$: SGS pressure-diatation\\
$\rho$: Density\\
$\sigma_{ij}$: Components of the SGS tensor\\
$\tau_{ij}:$ Shear-stress tensor\\

\subsection*{Subscripts}

$j$: Relative to the jet \\ 
$\infty$: Relative to the freestream \\
$e$: Relative to interior domain \\
$f$: Relative to boundary face \\
$sgs$: subgrid property


\end{multicols}

%% file: sources/theoretical/theoretical.tex
\section{Large Eddy Simulation Formulation}

The numerical strategy used in the present study 
is based on the System I filtered compressible 
Navier-Stokes equations \cite{Vreman1995} formulated 
as
\begin{eqnarray}
\displaystyle \frac{\partial \overline{\rho} }{\partial t} 
+ \frac{\partial}{\partial x_{j}} 
\left( \overline{\rho} \widetilde{  u_{j} } \right) 
&=& 0 \, \mbox{,}\\
\displaystyle \frac{\partial}{\partial t} 
\left( \overline{ \rho } \widetilde{ u_{i} } \right) 
+ \frac{\partial}{\partial x_{j}} \left( \overline{ \rho } 
\widetilde{ u_{i} } \widetilde{ u_{j} } \right)
+ \frac{\partial \overline{p}}{\partial x_{i}} 
- \frac{\partial \check{\tau}_{ij}}{\partial x_{j}} 
&=& -\frac{\partial}{\partial x_{j}} \left[ \sigma_{ij} 
- \left( \overline{\tau_{ij}} - \check{\tau}_{ij}  \right) \right] 
\, \mbox{,} \label{eq:les_qdm}\\ 
\displaystyle \frac{\partial \overline{\rho} \check{E}}{\partial t} 
+ \frac{\partial \left( \overline{\rho} \check{E} 
+ \overline{p} \right)\widetilde{u_{j}}}{\partial x_{j}} 
- \frac{\partial \check{\tau}_{ij} \widetilde{u_{i}} }{\partial x_{j}} 
+ \frac{\partial \check{q}_{j}}{\partial x_{j}} 
&=& -B_{1}-B_{2}-B_{3}+B_{4}+B_{5}+B_{6}-B_{7} \, \mbox{,}
\label{eq:les_energy}
\end{eqnarray}
in which $t$ and $x_{i}$ are independent variables 
representing time and spatial coordinates of a 
Cartesian coordinate system $\textbf{x}$, respectively. 
The components of the velocity vector $\textbf{u}$ 
are written as $u_{i}$, and $i=1,2,3$. Density, pressure 
and total energy per mass unit are denoted by $\rho$, 
$p$ and $E$, respectively. $\overline{\left( \cdot \right)}$ 
stands for filtered properties and $\tilde{\left( \cdot \right)}$ 
stands for Favre averaged properties. The filtered heat flux, 
$\check{q}_{j}$, is given by 
\begin{equation}
	\check{q}_{j} = \left[ \kappa ( \tilde{T} ) \right]
	\frac{\partial \tilde{T}}{\partial x_{j}} \, \mbox{,}
\end{equation} 
where $\tilde{T}$ is the averaged static temperature and 
$\kappa$ is the thermal conductivity, which 
can by expressed by
\begin{equation}
	\kappa ( T ) = \left[ \mu (T) \right] 
	\frac{ C_{p}}{Pr} \, \mbox{.}
\end{equation}
The thermal conductivity is a function of the specific heat 
at constant pressure, $Cp$, of the Prandtl number, $Pr$, 
which is equal to $0.72$ for air, and of the dynamic viscosity, 
$\mu$. The last is calculated using the Sutherland Law,
\begin{eqnarray}
\mu (T) = \mu_{\infty} 
\left( \frac{T}{T_{\infty}} \right)^{\frac{3}{2}} 
\frac{T_{0}+S_{1}}{T+S_{1}} &
\mbox{with} \: S_{1} = 110.4K \, \mbox{.}
\label{eq:sutherland}
\end{eqnarray}
In the present work, $\check{\tau}_{ij}$ is calculated
using the Stokes hypothesis for Newtonian fluids, 
\begin{equation}
	\check{\tau}_{ij} = 2 \left[ \mu (\tilde{T}) \right] 
	\left( \check{S}_{ij} - \frac{1}{3} \delta_{ij} \check{S}_{kk} 
\right) \, \mbox{,} 
\end{equation}
in which $\check{S}_{ij}$, components of rate-of-strain tensor, is given by
\begin{equation}
\check{S}_{ij} = \frac{1}{2} 
\left( \frac{\partial \widetilde{ u_{i} } }{\partial x_{j}} 
+ \frac{\partial \widetilde{ u_{j} } }{\partial x_{i}} \right) 
\, \mbox{.}
\end{equation}
In order to close the system of equations the density, the static 
pressure and the static temperature are correlated by the equation 
of state given by
\begin{equation}
	\overline{p} = \overline{\rho} R \tilde{T} \, \mbox{,}
\end{equation}
where $R$ is the gas constant, written as
\begin{equation}
R = C_{p} - C_{v} \, \mbox{,}
\end{equation}
and $C_{v}$ is the specif heat at constant volume. The 
System I formulation \cite{Vreman1995} neglects the double
correlation term of the total energy per mass unity and
writes
\begin{equation}
	\overline{\rho E} = \overline{\rho} \check{E} = 
	\frac{\overline{p}}{\gamma - 1} 
	+ \frac{1}{2} \rho \tilde{u}_{i} \tilde{u}_{i} \, \mbox{,} 
\end{equation}
in which $\gamma$ is the specif heat ratio written as $\gamma = Cp/Cv$.
The components of the SGS stress tensor, $\sigma$, 
are given by
\begin{equation}
\sigma_{ij} = \overline{\rho} \left( \widetilde{u_{i}u_{j}} 
- \widetilde{u_{i}} \widetilde{u_{j}} \right) 
\, \mbox{,}
\end{equation}
and the SGS $B_{i}$ terms are written as
\begin{eqnarray}
B_{1} = \frac{1}{(\gamma - 1)} 
\frac{\partial}{\partial x_{j}} \left( \overline{p u_{j}} 
- \overline{p} \widetilde{u_{j}} \right) 
= \frac{\partial C_{v} \mathcal{Q}_{j}}{\partial x_{j}} \, \mbox{,} \\
B_{2} = \overline{p \frac{\partial u_{k}}{\partial x_{k}} } 
- \overline{p} \frac{\partial \widetilde{u_{k}}}{\partial x_{k}} 
= \Pi_{dil} \, \mbox{,}\\
B_{3} = \frac{\partial}{\partial x_{j}} 
\left( \sigma_{kj} \widetilde{u_{k}} \right) \, \mbox{,} \\
B_{4} = \sigma_{kj} 
\frac{\partial}{\partial x_{j}} \widetilde{u_{k}} \, \mbox{,} \\
B_{5} = \overline{\tau_{kj} \frac{\partial}{\partial x_{j}} u_{k}} 
- \overline{\tau_{ij}} \frac{\partial}{\partial x_{j}}\widetilde{u_{k}} 
= \mathcal{E} \, \mbox{,} \\
B_{6} = \frac{\partial}{\partial x_{j}} 
\left( \overline{\tau_{ij}}\widetilde{u_{i}} 
- \check{\tau}_{ij}\widetilde{u_{i}} \right) 
= \frac{\partial \mathcal{D}}{\partial x_{j}} 
\, \mbox{,} \\
B_{7} = \frac{\partial}{\partial x_{j}} \left( \overline{q_{j}} 
- \check{q}_{j} \right) \, \mbox{,}
\end{eqnarray}
where $\mathcal{Q}_{j}$ is the SGS temperature flux, written as
\begin{equation}
	\mathcal{Q}_{j} = \overline{\rho} \left( \widetilde{u_{j} T} 
	- \widetilde{u_{j}} \widetilde{T} \right) \, \mbox{,}
\label{eq:sgs_temp_flux}
\end{equation}
$\Pi_{dil}$ is the SGS pressure-dilatation term,  
$\mathcal{E}$ is the SGS viscous dissipation and 
$\mathcal{D}$ is the SGS viscous diffusion term.

%% file: sources/turb_mod/turb_mod.tex
\section{Subgrid Scale Modeling} \label{ch:turb_mod}

The SGS terms presented in Eqs.\ \eqref{eq:les_qdm} and
\eqref{eq:les_energy} cannot be directly calculated. 
Subgrid scale models are necessary in order to close 
the LES set of equations. The closure models presented 
here are based on homogeneous turbulence theory, 
which is usually developed in spectral space as an 
atempt to quantify the interaction between the different 
scales of turbulence.

\subsection{Subgrid Scale Viscosity}

The SGS stress tensor is here calculated using
the concept that the forward energy cascade is analogous 
to the molecular mechanisms represented by the molecular 
viscosity whose mathematical structure follows the Boussinesq 
hypothesis \cite{Boussinesq1877} to explicitly introduce 
the subgrid scale viscosity, $\mu_{sgs}$,
\begin{equation}
\sigma_{ij}^{d} = \sigma_{ij} - \frac{1}{3} \delta_{ij} \sigma_{kk} = 
-2 \mu_{sgs} \left( \check{S}_{ij} - \frac{1}{3} \check{S}_{kk} \right) \, \mbox{.}
\end{equation}
The isotropic portion of the SGS stress tensor is 
generally weak with respect to the thermodynamic 
pressure \cite{Erlebacher92}. Therefore, $\sigma_{kk}$
is neglected in the present paper and $\sigma_{ij}$ 
is written as
\begin{equation}
\sigma_{ij} = -2 \mu_{sgs} \left( \check{S}_{ij} 
- \frac{1}{3} \check{S}_{kk} \right) \, \mbox{.}
\end{equation}
The Smagorinsky model \cite{Smagorinsky63, Lilly65} is used to 
calculate the SGS viscosity and it is given by 
\begin{equation}
	\mu_{sgs} = \overline{\rho} \left( C_{s} \Delta \right)^{2}  
\left( 2 \check{S}_{ij} \check{S}_{ij} \right)^{\frac{1}{2}}
\, \mbox{,}
\end{equation}
and $C_{s}$ is the Smagorinsky constant. Several 
attempts can be found in the literature regarding the evaluation
of the Smagorinsky constant. The value of this constant is adjusted to 
improve the results of different flow configurations. In pratical terms, 
the Smagorinsky constant assumes value ranging from 0.1 to 0.2 depending on the 
flow. The present work uses $C_{s}=0.18$ as suggested by Lilly
\cite{Lilly67}.

\subsection{Subgrid Scale Modeling of the Energy Equation}

Some of the SGS terms within the System I formulation 
\cite{Vreman1995} are not directly computable. Therefore, modeling 
is applied in order to close the system. Several papers in the literature
\cite{Garnier09,Erlebacher92,moin91,Martin00,lenormand2000,coleman1995} 
evaluate the relevance of the SGS terms on different applications. 
The LES community commonly uses the hypothesis of Eidson 
\cite{eidson85}, which assumes that the energy transfer from the 
resolved scales is proportional to the gradient of the resolved 
temperature \cite{Garnier09}. The proportionality coefficient 
is the subgrid scale conductivity, $\kappa_{sgs}$, which is 
linked to the subgrid scale viscosity through the relation 
\begin{equation}
	\kappa_{sgs} = \frac{\overline{\rho}\, \nu_{sgs}\, C_{p}}{{Pr}_{sgs}} 
	\, \mbox{,}
\end{equation}
in which $Pr_{sgs}$ is the SGS Prandtl number. The eddy-damped 
quasi-normal Markovian (EDQNM) theory \cite{lesieur08} considers 
$Pr_{sgs}=0.6$. In the present work, the concept of SGS conductivity, 
$\kappa_{sgs}$, is used in order to model $B_{1} + B_{2}$, {\em i.e.}, 
\begin{equation}
	B_{1} + B_{2} = -\frac{\partial}{\partial x_{j}} 
	\left( \kappa_{sgs} \frac{\partial \tilde{T}}{\partial x_{j}} \right) 
	\, \mbox{,}
\end{equation}
Following the work of Larcheveque \cite{larcheveque03}, $B_{4}$, $B_{5}$, 
$B_{6}$ and $B_{7}$ are neglected in the present paper.

\subsection{Shear-Stress Tensor Modeling}

The difference between the shear-stress tensors at the right-hand side of the 
momentum equations is neglected in the present work,
\begin{equation}
    \overline{\tau_{ij}} - \check{\tau}_{ij} = 0
\end{equation}

%% file: sources/coord/coord.tex
\section{Transformation of Coordinates}

The legacy code formulation \cite{BIGA02} was originally written in the a general curvilinear coordinate 
system in order to facilitate the implementation and add more generality for the CFD tool. This approach is 
kept in the present LES solver for the simulation of compressible jet flows. Hence, the filtered Navier-Stokes 
equations can be written in strong conservation form for a 3-D general curvilinear coordinate system as
\begin{equation}
	\frac{\partial \hat{Q}}{\partial \mathcal{T}} 
	+ \frac{\partial }{\partial \xi}\left(\hat{\mathbf{E}_{e}}-\hat{\mathbf{E}_{v}}\right) 
	+ \frac{\partial}{\partial \eta}\left(\hat{\mathbf{F}_{e}}-\hat{\mathbf{F}_{v}}\right)
	+ \frac{\partial}{\partial \zeta}\left(\hat{\mathbf{G}_{e}}-\hat{\mathbf{G}_{v}}\right) 
	= 0 \, \mbox{.}
	\label{eq:vec-LES}
\end{equation}
In the present work, the chosen general coordinate transformation is given by
\begin{eqnarray}
	\mathcal{T} & = & t \, \mbox{,} \nonumber\\
	\xi & = & \xi \left(x,y,z,t \right)  \, \mbox{,} \nonumber\\
	\eta & = & \eta \left(x,y,z,t \right)  \, \mbox{,} \\
	\zeta & = & \zeta \left(x,y,z,t \right)\ \, \mbox{.} \nonumber
\end{eqnarray}
In the jet flow configuration, $\xi$ is the axial jet flow direction, $\eta$ is 
the radial direction and $\zeta$ is the azimuthal direction. The vector of
conserved properties is written as
\begin{equation}
	\hat{Q} = J^{-1} \left[ \overline{\rho} \quad \overline{\rho}\tilde{u} \quad 
	\overline{\rho}\tilde{v} \quad \overline{\rho}\tilde{w} \quad \overline{e} \right]^{T} 
	\quad \mbox{,}
	\label{eq:hat_Q_vec}
\end{equation}
where the Jacobian of the transformation, $J$, is given by
\begin{equation}
	J = \left( x_{\xi} y_{\eta} z_{\zeta} + x_{\eta}y_{\zeta}z_{\xi} +
	           x_{\zeta} y_{\xi} z_{\eta} - x_{\xi}y_{\zeta}z_{\eta} -
			   x_{\eta} y_{\xi} z_{\zeta} - x_{\zeta}y_{\eta}z_{\xi} 
	    \right)^{-1} \, \mbox{,}
\end{equation}
and
\begin{eqnarray}
	\displaystyle x_{\xi}   = \frac{\partial x}{\partial \xi}  \, \mbox{,} & 
	\displaystyle x_{\eta}  = \frac{\partial x}{\partial \eta} \, \mbox{,} & 
	\displaystyle x_{\zeta} = \frac{\partial x}{\partial \zeta}\, \mbox{,} \nonumber \\
	\displaystyle y_{\xi}   = \frac{\partial y}{\partial \xi}  \, \mbox{,} & 
	\displaystyle y_{\eta}  = \frac{\partial y}{\partial \eta} \, \mbox{,} & 
	\displaystyle y_{\zeta} = \frac{\partial y}{\partial \zeta}\, \mbox{,} \\
	\displaystyle z_{\xi}   = \frac{\partial z}{\partial \xi}  \, \mbox{,} & 
	\displaystyle z_{\eta}  = \frac{\partial z}{\partial \eta} \, \mbox{,} & 
	\displaystyle z_{\zeta} = \frac{\partial z}{\partial \zeta}\, \mbox{.} \nonumber
\end{eqnarray}

The inviscid flux vectors, $\hat{\mathbf{E}}_{e}$, $\hat{\mathbf{F}}_{e}$ and 
$\hat{\mathbf{G}}_{e}$, are given by
{\small
\begin{eqnarray}
	\hat{\mathbf{E}}_{e} = J^{-1} \left\{\begin{array}{c}
		\overline{\rho} U \\
		\overline{\rho}\tilde{u} U + \overline{p} \xi_{x} \\
		\overline{\rho}\tilde{v} U + \overline{p} \xi_{y} \\
		\overline{\rho}\tilde{w} U + \overline{p} \xi_{z} \\
		\left( \overline{e} + \overline{p} \right) U - \overline{p} \xi_{t}
\end{array}\right\} \, \mbox{,} &
%
	\hat{\mathbf{F}}_{e} = J^{-1} \left\{\begin{array}{c}
		\overline{\rho} V \\
		\overline{\rho}\tilde{u} V + \overline{p} \eta_{x} \\
		\overline{\rho}\tilde{v} V + \overline{p} \eta_{y} \\
		\overline{\rho}\tilde{w} V + \overline{p} \eta_{z} \\
		\left( \overline{e} + \overline{p} \right) V - \overline{p} \eta_{t}
\end{array}\right\} \, \mbox{,} &
%
	\hat{\mathbf{G}}_{e} = J^{-1} \left\{\begin{array}{c}
		\overline{\rho} W \\
		\overline{\rho}\tilde{u} W + \overline{p} \zeta_{x} \\
		\overline{\rho}\tilde{v} W + \overline{p} \zeta_{y} \\
		\overline{\rho}\tilde{w} W + \overline{p} \zeta_{z} \\
		\left( \overline{e} + \overline{p} \right) W - \overline{p} \zeta_{t}
	\end{array}\right\} \, \mbox{.}
	\label{eq:hat-flux-G}
\end{eqnarray}
}
The contravariant velocity components, $U$, $V$ and $W$, are calculated as
\begin{eqnarray}
  U = \xi_{t} + \xi_{x}\overline{u} + \xi_{y}\overline{v} + \xi_{z}\overline{w} 
  \, \mbox{,} \nonumber \\
  V = \eta_{t} + \eta_{x}\overline{u} + \eta_{y}\overline{v} + \eta_{z}\overline{w} 
  \, \mbox{,} \\
  W = \zeta_{t} + \zeta_{x}\overline{u} + \zeta_{y}\overline{v} + \zeta_{z}\overline{w} 
  \, \mbox{.} \nonumber
  \label{eq:vel_contrv}
\end{eqnarray}
The metric terms are given by
\begin{eqnarray}
	\xi_{x} = J \left( y_{\eta}z_{\zeta} - y_{\zeta}z_{\eta} \right) \, \mbox{,} & 
	\xi_{y} = J \left( z_{\eta}x_{\zeta} - z_{\zeta}x_{\eta} \right) \, \mbox{,} & 
	\xi_{z} = J \left( x_{\eta}y_{\zeta} - x_{\zeta}y_{\eta} \right) \, \mbox{,} \nonumber \\
	\eta_{x} = J \left( y_{\eta}z_{\xi} - y_{\xi}z_{\eta} \right) \, \mbox{,} & 
	\eta_{y} = J \left( z_{\eta}x_{\xi} - z_{\xi}x_{\eta} \right) \, \mbox{,} & 
	\eta_{z} = J \left( x_{\eta}y_{\xi} - x_{\xi}y_{\eta} \right) \, \mbox{,} \\
	\zeta_{x} = J \left( y_{\xi}z_{\eta} - y_{\eta}z_{\xi} \right) \, \mbox{,} & 
	\zeta_{y} = J \left( z_{\xi}x_{\eta} - z_{\eta}x_{\xi} \right) \, \mbox{,} & 
	\zeta_{z} = J \left( x_{\xi}y_{\eta} - x_{\eta}y_{\xi} \right) \, \mbox{,} \nonumber \\
	\xi_{t} = -x_{\mathcal{T}}\xi_{x} - y_{\mathcal{T}}\xi_{y} - z_{\mathcal{T}}\xi_{z} \, \mbox{,} & 
	\eta_{t} = -x_{\mathcal{T}}\eta_{x} - y_{\mathcal{T}}\eta_{y} - z_{\mathcal{T}}\eta_{z} \, \mbox{,} & 
	\zeta_{t} = -x_{\mathcal{T}}\zeta_{x} - y_{\mathcal{T}}\zeta_{y} - z_{\mathcal{T}}\zeta_{z} \, \mbox{.} 
	\nonumber
\end{eqnarray}

The viscous flux vectors, $\hat{\mathbf{E}}_{v}$, $\hat{\mathbf{F}}_{v}$ and 
$\hat{\mathbf{G}}_{v}$, are written as
\begin{equation}
	\hat{\mathbf{E}}_{v} = J^{-1} \left\{\begin{array}{c}
		0 \\
		\xi_{x}\hat{\tau}_{xx} +  \xi_{y}\hat{\tau}_{xy} + \xi_{z}\hat{\tau}_{xz} \\
		\xi_{x}\hat{\tau}_{xy} +  \xi_{y}\hat{\tau}_{yy} + \xi_{z}\hat{\tau}_{yz} \\
		\xi_{x}\hat{\tau}_{xz} +  \xi_{y}\hat{\tau}_{yz} + \xi_{z}\hat{\tau}_{zz} \\
		\xi_{x}{\beta}_{x} +  \xi_{y}{\beta}_{y} + \xi_{z}{\beta}_{z} 
	\end{array}\right\} \, \mbox{,}
	\label{eq:hat-flux-Ev}
\end{equation}
\begin{equation}
	\hat{\mathbf{F}}_{v} = J^{-1} \left\{\begin{array}{c}
		0 \\
		\eta_{x}\hat{\tau}_{xx} +  \eta_{y}\hat{\tau}_{xy} + \eta_{z}\hat{\tau}_{xz} \\
		\eta_{x}\hat{\tau}_{xy} +  \eta_{y}\hat{\tau}_{yy} + \eta_{z}\hat{\tau}_{yz} \\
		\eta_{x}\hat{\tau}_{xz} +  \eta_{y}\hat{\tau}_{yz} + \eta_{z}\hat{\tau}_{zz} \\
		\eta_{x}{\beta}_{x} +  \eta_{y}{\beta}_{y} + \eta_{z}{\beta}_{z} 
	\end{array}\right\} \, \mbox{,}
	\label{eq:hat-flux-Fv}
\end{equation}
\begin{equation}
	\hat{\mathbf{G}}_{v} = J^{-1} \left\{\begin{array}{c}
		0 \\
		\zeta_{x}\hat{\tau}_{xx} +  \zeta_{y}\hat{\tau}_{xy} + \zeta_{z}\hat{\tau}_{xz} \\
		\zeta_{x}\hat{\tau}_{xy} +  \zeta_{y}\hat{\tau}_{yy} + \zeta_{z}\hat{\tau}_{yz} \\
		\zeta_{x}\hat{\tau}_{xz} +  \zeta_{y}\hat{\tau}_{yz} + \zeta_{z}\hat{\tau}_{zz} \\
		\zeta_{x}{\beta}_{x} +  \zeta_{y}{\beta}_{y} + \zeta_{z}{\beta}_{z} 
	\end{array}\right\} \, \mbox{,}
	\label{eq:hat-flux-Gv}
\end{equation}
where $\beta_{x}$, $\beta_{y}$ and $\beta_{z}$ are defined as
\begin{eqnarray}
	\beta_{x} = \hat{\tau}_{xx}\tilde{u} + \hat{\tau}_{xy}\tilde{v} +
	\hat{\tau}_{xz}\tilde{w} - \overline{q}_{x} \, \mbox{,} \nonumber \\
	\beta_{y} = \hat{\tau}_{xy}\tilde{u} + \hat{\tau}_{yy}\tilde{v} +
	\hat{\tau}_{yz}\tilde{w} - \overline{q}_{y} \, \mbox{,} \\
	\beta_{z} = \hat{\tau}_{xz}\tilde{u} + \hat{\tau}_{yz}\tilde{v} +
	\hat{\tau}_{zz}\tilde{w} - \overline{q}_{z} \mbox{.} \nonumber
\end{eqnarray}
and $\hat{\tau}_{ij}$ is given by
\begin{equation}
	\hat{\tau}_{ij} = 2 \left( \mu + \mu_{sgs}\right) 
	\left( \check{S}_{ij} - \frac{1}{3} \check{S}_{kk} \right) \, \mbox{.}
\end{equation}

%% file: sources/dimlss/dimlss.tex
\section{Dimensionless Formulation}

A convenient nondimensionalization is necessary in to order to achieve a consistent 
implementation of the governing equations of motion. Dimensionless formulation 
yields to a more general numerical tool. There is no need to change the formulation 
for each configuration intended to be simulated. Moreover, dimensionless formulation 
scales all the necessary properties to the same order of magnitude which is a 
computational advantage \cite{BIGA02}. Dimensionless variables are presented in the 
present section in order perform the nondimensionalization of Eq.\ 
\eqref{eq:vec-LES}

The dimensionless time, $\underline{\mathcal{T}}$, is written as function of the 
speed of sound of the jet at the inlet, $a_{j}$, and of a reference lenght, $l$,
\begin{equation}
	\underline{\mathcal{T}} = \mathcal{T} \frac{a_{j}}{l} \, \mbox{.}
	\label{eq:non-dim-time}
\end{equation}
In the present work $l$ represents the jet entrance diameter $d$. This reference lenght is 
also applied to write the dimensionless length, 
\begin{equation}
	\underline{l} = \frac{l}{d} \, \mbox{.}
	\label{eq:non-dim-lengh}
\end{equation}
The dimensionless velocity components are obtained using the speed of sound of the 
jet at the inlet,
\begin{equation}
	\underline{\textbf{u}} = \frac{\textbf{u}}{a_{j}} \, \mbox{.}
	\label{eq:non-dim-vel}
\end{equation}
Dimensionless pressure and energy are calculated using density and speed of the sound
of the jet at the inlet as
\begin{equation}
	\underline{p} = \frac{p}{\rho_{j}a_{j}^{2}} \, \mbox{,}
	\label{eq:non-dim-press}
\end{equation}
\begin{equation}
	\underline{E} = \frac{E}{\rho_{j}a_{j}^{2}} \, \mbox{.}
	\label{eq:non-dim-energy}
\end{equation}
Dimensionless density, $\underline{\rho}$, temperature, $\underline{T}$ and 
viscosity, $\underline{\mu}$, are calculated using freestream properties
\begin{equation}
	\underline{\rho} = \frac{\rho}{\rho_{j}} \, \mbox{.}
	\label{eq:non-dim-rho}
\end{equation}

One can use the dimensionless properties described above in order to write the 
dimensionless form of the RANS equations as
\begin{equation}
	\frac{\partial \underline{Q}}{\partial \mathcal{T}} + 
	\frac{\partial \underline{\mathbf{E}}_{e}}{\partial \xi} +
	\frac{\partial \underline{\mathbf{F}}_{e}}{\partial \eta} + 
	\frac{\partial \underline{\mathbf{G}}_{e}}{\partial \zeta} =
	\frac{1}{Re} \left( \frac{\partial \underline{\mathbf{E}}_{v}}{\partial \xi} 
	+ \frac{\partial \underline{\mathbf{F}}_{v}}{\partial \eta} 
	+ \frac{\partial \underline{\mathbf{G}}_{v}}{\partial \zeta} \right)	\, \mbox{,}
	\label{eq:vec-underline-split-RANS}
\end{equation}
where the underlined terms are calculated using dimensionless properties and the Reynolds number
is based on the jet properties such as the speed of sound, $a_{j}$, density, $\rho_{j}$, viscosity,
$\mu_{j}$ and the jet entrance diameter, $d$,
\begin{equation}
	Re = \frac{\rho_{j}a_{j}d}{\mu_{j}} \, \mbox{.}
\end{equation}

%% file: sources/numerical/numerical.tex
\section{Numerical Formulation}

The governing equations previously described are discretized in a 
structured finite difference context for general curvilinear 
coordinate system \cite{BIGA02}. The numerical flux is calculated 
through a central difference scheme with the explicit addition 
of the anisotropic scalar artificial dissipation of Turkel and Vatsa
\cite{Turkel_Vatsa_1994}. The time integration is performed by an 
explicit, 2nd-order, 5-stage Runge-Kutta scheme 
\cite{jameson_mavriplis_86, Jameson81}.  Conserved properties
and artificial dissipation terms are properly treated near boundaries in order
to assure the physical correctness of the numerical formulation. 
 
\subsection{Spatial Discretization}

For the sake of simplicity the formulation discussed in the present section
is no longer written using bars. However, the reader should notice that the 
equations are dimensionless and filtered. The Navier-Stokes equations, 
presented in Eq.\ \eqref{eq:vec-underline-split-RANS}, are discretized in 
space in a finite difference fashion and, then, rewritten as
\begin{equation}
	\left(\frac{\partial Q}{\partial \mathcal{T}}\right)_{i,j,k} \  
	= \  - RHS_{i,j,k} \, \mbox{,}	
	\label{eq:spatial_discret}
\end{equation}
where $RHS$ is the right hand side of the equation and it is written as function of 
the numerical flux vectors at the interfaces between grid points,
\begin{eqnarray}
	{RHS}_{i,j,k} & = & 
	\frac{1}{\Delta \xi} \left( 
	{\mathbf{E}_{e}}_{(i+\frac{1}{2},j,k)} - {\mathbf{E}_{e}}_{(i-\frac{1}{2},j,k)} - 
	{\mathbf{E}_{v}}_{(i+\frac{1}{2},j,k)} + {\mathbf{E}_{v}}_{(i-\frac{1}{2},j,k)} 
	\right) \nonumber \\
	& & \frac{1}{\Delta \eta} \left( 
	{\mathbf{F}_{e}}_{(i,j+\frac{1}{2},k)} - {\mathbf{F}_{e}}_{(i,j-\frac{1}{2},k)} - 
	{\mathbf{F}_{v}}_{(i,j+\frac{1}{2},k)} + {\mathbf{F}_{v}}_{(i,j-\frac{1}{2},k)} 
	\right) \\
	& & \frac{1}{\Delta \zeta} \left( 
	{\mathbf{G}_{e}}_{(i,j,k+\frac{1}{2})} - {\mathbf{G}_{e}}_{(i,j,k-\frac{1}{2})} - 
	{\mathbf{G}_{v}}_{(i,j,k+\frac{1}{2})} + {\mathbf{G}_{v}}_{(i,j,k-\frac{1}{2})} 
	\right) \, \mbox{.} \nonumber
\end{eqnarray}
For the general curvilinear coordinate case 
$\Delta \xi = \Delta \eta = \Delta \zeta = 1$. The anisotropic scalar 
artificial dissipation method of Turkel and Vatsa \cite{Turkel_Vatsa_1994}
is implemented through the modification of the inviscid flux vectors, 
$\mathbf{E}_{e}$, $\mathbf{F}_{e}$ and $\mathbf{G}_{e}$. The numerical scheme 
is nonlinear and allows the selection between artificial dissipation terms of 
second and fourth differences, which is very important for capturing discontinuities 
in the flow. The numerical fluxes are calculated at interfaces in order to reduce 
the size of the calculation cell and, therefore, facilitate the implementation 
of second derivatives since the the concept of numerical fluxes vectors is 
used for flux differencing. Only internal interfaces receive the corresponding 
artificial dissipation terms, and differences of the viscous flux vectors 
use two neighboring points of the interface. 

The inviscid flux vectors, with the addition of the artificial dissipation
contribution, can be written as
\begin{eqnarray}
	{\mathbf{E}_{e}}_{(i \pm \frac{1}{2},j,k)} 
	= \frac{1}{2} \left( {\mathbf{E}_{e}}_{(i,j,k)} + {\mathbf{E}_{e}}_{(i \pm 1,j,k)} \right)
	- J^{-1} \mathbf{d}_{(i \pm \frac{1}{2},j,k)} \, \mbox{,} \nonumber \\
	{\mathbf{F}_{e}}_{(i,j\pm \frac{1}{2},k)} 
	= \frac{1}{2} \left( {\mathbf{F}_{e}}_{(i,j,k)} + {\mathbf{F}_{e}}_{(i,j \pm 1,k)} \right)
	- J^{-1} \mathbf{d}_{(i,j \pm \frac{1}{2},k)} \, \mbox{,} \label{eq:inv_flux_vec}\\
	{\mathbf{G}_{e}}_{(i,j,k\pm \frac{1}{2})} 
	= \frac{1}{2} \left( {\mathbf{G}_{e}}_{(i,j,k)} + {\mathbf{G}_{e}}_{(i,j,k \pm 1)} \right)
	- J^{-1} \mathbf{d}_{(i,j,k \pm \frac{1}{2})} \, \mbox{,} \nonumber
\end{eqnarray}
in which the $\mathbf{d}_{(i\pm 1,j,k)}$,$\mathbf{d}_{(i,j\pm 1,k)}$ and $\mathbf{d}_{(i,j,k\pm 1)}$ terms
are the Turkel and Vatsa \cite{Turkel_Vatsa_1994} artificial dissipation terms
in the $i$, $j$, and $k$ directions respectively. The scaling of the artificial
dissipation operator in each coordinate direction is weighted by its own spectral 
radius of the corresponding flux Jacobian matrix, which gives the non-isotropic 
characteristics of the method \cite{BIGA02}. The artificial dissipation contribution
in the $\xi$ direction is given by
\begin{eqnarray}
	\mathbf{d}_{(i + \frac{1}{2},j,k)} & = & 
	\lambda_{(i + \frac{1}{2},j,k)} \left[ \epsilon_{(i + \frac{1}{2},j,k)}^{(2)}
	\left( \mathcal{W}_{(i+1,j,k)} - \mathcal{W}_{(i,j,k)} \right) \right. \label{eq:dissip_term}\\
	& & \epsilon_{(i + \frac{1}{2},j,k)}^{(4)} \left( \mathcal{W}_{(i+2,j,k)} 
	- 3 \mathcal{W}_{(i+1,j,k)} + 3 \mathcal{W}_{(i,j,k)} 
	- \mathcal{W}_{(i-1,j,k)} \right) \left. \right] \, \mbox{,} \nonumber
\end{eqnarray}
in which
\begin{eqnarray}
	\epsilon_{(i + \frac{1}{2},j,k)}^{(2)} & = &
	k^{(2)} \mbox{max} \left( \nu_{(i+1,j,k)}^{d}, 
	\nu_{(i,j,k)}^{d} \right) \, \mbox{,} \label{eq:eps_2_dissip} \\
	\epsilon_{(i + \frac{1}{2},j,k)}^{(4)} & = &
	\mbox{max} \left[ 0, k^{(4)} - \epsilon_{(i + \frac{1}{2},j,k)}^{(2)} \right] 
	\, \mbox{.} \label{eq:eps_4_dissip}
\end{eqnarray}
The original article \cite{Turkel_Vatsa_1994} recomends using $k^{(2)}=0.25$ and 
$k^{(4)}=0.016$ for the dissipation artificial constants. The pressure 
gradient sensor, $\nu_{(i,j,k)}^{d}$, for the $\xi$ direction is written as
\begin{equation}
	\nu_{(i,j,k)}^{d} = \frac{|p_{(i+1,j,k)} - 2 p_{(i,j,k)} + p_{(i-1,j,k)}|}
	                          {p_{(i+1,j,k)} - 2 p_{(i,j,k)} + p_{(i-1,j,k)}} 
	\, \mbox{.}
\label{eq:p_grad_sensor}
\end{equation}
The $\mathcal{W}$ vector from Eq.\ \eqref{eq:dissip_term} is calculated as a function of the
conserved variable vector, $\hat{Q}$, written in Eq.\ \eqref{eq:hat_Q_vec}.
The formulation intends to keep the total enthalpy constant in the final converged 
solution, which is the correct result for the Navier-Stokes equations with 
$Re \rightarrow \infty$. This approach is also valid for the viscous formulation 
because the dissipation terms are added to the inviscid flux terms, in which they 
are really necessary to avoid nonlinear instabilities of the numerical formulation. 
The $\mathcal{W}$ vector is given by
\begin{equation}
	\mathcal{W} = \hat{Q} + \left[0 \,\, 0 \,\, 0 \,\, 0 \,\, p \right]^{T} \, \mbox{.}
	\label{eq:W_dissip}
\end{equation}
The spectral radius-based scaling factor, $\lambda$, for the $i-\mbox{th}$ 
direction is written
\begin{equation}
	\lambda_{(i+\frac{1}{2},j,k)} = \frac{1}{2} \left[ 
	\left( \overline{\lambda_{\xi}}\right)_{(i,j,k)} + 
	\left( \overline{\lambda_{\xi}}\right)_{(i+1,j,k)}
	\right] \, \mbox{,} 
\end{equation}
where
\begin{equation}
    \overline{\lambda_{\xi}}_{(i,j,k)} = \lambda_{\xi} \left[ 1 + 
	\left(\frac{\lambda_{\eta}}{\lambda_{\xi}} \right)^{0.5} + 
	\left(\frac{\lambda_{\zeta}}{\lambda_{\xi}} \right)^{0.5} \right] 
	\, \mbox{.}
\end{equation}
The spectral radii, $\lambda_{\xi}$, $\lambda_{\eta}$ and $\lambda_{\zeta}$ are given
by
\begin{eqnarray}
	\lambda_{\xi} & = & 
	|U| + a \sqrt{\xi_{x}^{2} + \eta_{y}^{2} + \zeta_{z}^{2}} 
	\, \mbox{,} \nonumber \\
	\lambda_{\xi} & = & 
	|V| + a \sqrt{\xi_{x}^{2} + \eta_{y}^{2} + \zeta_{z}^{2}} 
	\, \mbox{,} \\
	\lambda_{\xi} & = & 
	|W| + a \sqrt{\xi_{x}^{2} + \eta_{y}^{2} + \zeta_{z}^{2}} 
	\, \mbox{,} \nonumber
\end{eqnarray}
in which, $U$, $V$ and $W$ are the contravariants velocities in the $\xi$, $\eta$
and $\zeta$, previously written in Eq.\ \eqref{eq:vel_contrv}, and $a$ is the local 
speed of sound, which can be written as
\begin{equation}
	a = \sqrt{\frac{\gamma p}{\rho}} \, \mbox{.}
\end{equation}
The calculation of artificial dissipation terms for the other coordinate directions
are completely similar and, therefore, they are not discussed in the present work.

\subsection{Time Marching Method}

The time marching method used in the present work is a 2nd-order, 5-step Runge-Kutta
scheme based on the work of Jameson \cite{Jameson81, jameson_mavriplis_86}. 
The time integration can be written as
\begin{equation}
	\begin{array}{ccccc}
	Q_{(i,jk,)}^{(0)} & = & Q_{(i,jk,)}^{(n)} \, \mbox{,} & & \\
	Q_{(i,jk,)}^{(l)} & = & Q_{(i,jk,)}^{(0)} -  
	& \alpha_{l} {\Delta t}_{(i,j,k)} {RHS}_{(i,j,k)}^{(l-1)} \, & 
	\,\,\,\, l = 1,2 \cdots 5, \\
	Q_{(i,jk,)}^{(n+1)} & = & Q_{(i,jk,)}^{(5)} \, \mbox{,} & &
	\end{array}
	\label{eq:localdt}
\end{equation}
in which $\Delta t$ is the time step and $n$ and $n+1$ indicate the property
values at the current and at the next time step, respectively. The literature
\cite{Jameson81, jameson_mavriplis_86} recommends 
\begin{equation}
	\begin{array}{ccccc}
		\alpha_{1} = \frac{1}{4} \,\mbox{,} & \alpha_{2} = \frac{1}{6} \,\mbox{,} &
		\alpha_{3} = \frac{3}{8} \,\mbox{,} & \alpha_{4} = \frac{1}{2} \,\mbox{,} & 
		\alpha_{5} = 1 \,\mbox{,} 
	\end{array}
\end{equation}
in order to improve the numerical stability of the time integration. The present
scheme is theoretically stable for $CFL \leq 2\sqrt{2}$, under a linear analysis
\cite{BIGA02}.

\section{Boundary Conditions} \label{sec:BC}

The original solver was created for two flow configurations, the 3D zero-pressure gradient
flat plate and the 3D axisymmetric flows. For the configuration of interest here, the 
types of boundary conditions include a combination of the boundary conditions coming from
the two original cases: entrance and exit conditions, centerline and far fields conditions.
Considering the new case as perfectly 3D, instead of the symmetric condition, a new 
periodic condition is implemented in the azimuthal direction.  

\subsection{Far Field Boundary}

Riemann invariants \cite{Long91} are used to implement far field boundary conditions.
They are derived from the characteristic relations for the Euler equations.
At the interface of the outer boundary, the following expressions apply
\begin{eqnarray}
	\mathbf{R}^{-} = {\mathbf{R}}_{\infty}^{-} & = & q_{n_\infty}-\frac{2}{\gamma-1}a_\infty\, \mbox{,}  \\
	\mathbf{R}^{+} = {\mathbf{R}}_{e}^{+} & = & q_{n_e}-\frac{2}{\gamma-1}a_e \, \mbox{,}
	\label{eq:R-farfield}
\end{eqnarray}
where $\infty$ and $e$ indexes stand for the property in the freestream and in the 
internal region, respectively. $q_n$ is the velocity component normal to the outer surface,
defined as
\begin{equation}
	q_n={\bf u} \cdot \vec{n} \, \mbox{,}
	\label{eq:qn-farfield}
\end{equation}
and $\vec{n}$ is the unit outward normal vector 
\begin{equation}
	\vec{n}=\frac{1}{\sqrt{\eta_{x}^2+\eta_{y}^2+\eta_{z}^2}}
	[\eta_x \ \eta_y \ \eta_z ]^T \, \mbox{.}
	\label{eq:norm-vec}
\end{equation}
Equation \eqref{eq:qn-farfield} assumes that the $\eta$ direction is pointing from the jet to the 
external boundary. Solving for $q_n$ and $a$, one can obtain
\begin{eqnarray}
	q_{n f} = \frac{\mathbf{R}^+ + \mathbf{R}^-}{2} \, \mbox{,} & \ & 
	a_f = \frac{\gamma-1}{4}(\mathbf{R}^+ - \mathbf{R}^-) \, \mbox{.}
	\label{eq: qn2-farfield}
\end{eqnarray}
The index $f$ is linked to the property at the boundary surface and will be used to update 
the solution at this boundary. For a subsonic exit boundary, $0<q_{n_e}/a_e<1$, the 
velocity components are derived from internal properties as
 \begin{eqnarray}
	 u_f&=&u_e+(q_{n f}-q_{n_e})\eta_x \, \mbox{,} \nonumber \\ 
	 v_f&=&v_e+(q_{n f}-q_{n_e})\eta_y \, \mbox{,} \\ 
	 w_f&=&w_e+(q_{n f}-q_{n_e})\eta_z \, \mbox{.} \nonumber
	 \label{eq:vel-farfield}
 \end{eqnarray}
Density and pressure properties are obtained by extrapolating the entropy from 
the adjacent grid node,
\begin{eqnarray}
	\rho_f = 
	\left(\frac{\rho_{e}^{\gamma}a_{f}^2}{\gamma p_e} \right)^{\frac{1}{\gamma-1}}
	\, \mbox{,} & \ &
	p_{f} = \frac{\rho_{f} a_{f}^2}{\gamma} \, \mbox{.} \nonumber
	 \label{eq:rhop-farfield}
\end{eqnarray}
For a subsonic entrance, $-1<q_{n_e}/a_e<0$, properties are obtained similarly 
from the freestream variables as
\begin{eqnarray}
	u_f&=&u_\infty+(q_{n f}-q_{n_\infty})\eta_x \, \mbox{,} \nonumber \\
	v_f&=&v_\infty+(q_{n f}-q_{n_\infty})\eta_y \, \mbox{,} \\
	w_f&=&w_\infty+(q_{n f}-q_{n_\infty})\eta_z \, \mbox{,} \nonumber
	\label{eq:vel2-farfield}
\end{eqnarray}
\begin{equation}
	\rho_f = 
	\left(\frac{\rho_{\infty}^{\gamma}a_{f}^2}{\gamma p_\infty} \right)^{\frac{1}{\gamma-1}}
	\, \mbox{.}
	\label{eq:rhop2-farfield}
\end{equation}
For a supersonic exit boundary, $q_{n_e}/a_e>1$, the properties are extrapolated 
from the interior of the domain as
\begin{eqnarray}
	\rho_f&=&\rho_e \, \mbox{,} \nonumber\\
	u_f&=&u_e \, \mbox{,} \nonumber\\
	v_f&=&v_e \, \mbox{,} \\
	w_f&=&w_e \, \mbox{,} \nonumber\\
	e_f&=&e_e \, \mbox{,} \nonumber   
	\label{eq:supso-farfield}
\end{eqnarray}
and for a supersonic entrance, $q_{n_e}/a_e<-1$, the properties are extrapolated 
from the freestream variables as
\begin{eqnarray}
	\rho_f&=&\rho_\infty \, \mbox{,}  \nonumber\\
	u_f&=&u_\infty \, \mbox{,}  \nonumber\\
	v_f&=&v_\infty \, \mbox{,} \\
	w_f&=&w_\infty \, \mbox{,} \nonumber\\
	e_f&=&e_\infty \, \mbox{.} \nonumber
	\label{eq:supso2-farfield}
\end{eqnarray}

\subsection{Entrance Boundary}

For a jet-like configuration, the entrance boundary is divided in two areas: the
jet and the area above it. The jet entrance boundary condition is implemented through 
the use of the 1-D characteristic relations for the 3-D Euler equations for a flat
velocity profile. The set of properties then determined is computed from within and 
from outside the computational domain. For the subsonic entrance, the $v$ and $w$ components
of the velocity are extrapolated by a zero-order extrapolation from inside the 
computational domain and the angle of flow entrance is assumed fixed. The rest of the properties 
are obtained as a function of the jet Mach number, which is a known variable. 
\begin{eqnarray}
	\left( u \right)_{1,j,k} & = & u_{j} \, \mbox{,} \nonumber \\
	\left( v \right)_{1,j,k} & = & \left( v \right)_{2,j,k} \,\mbox{,} \\
	\left( w \right)_{1,j,k} & = & \left( w \right)_{2,j,k} \, \mbox{.} \nonumber
	\label{eq:vel-entry}
\end{eqnarray}
The dimensionless total temperature and total pressure are defined with the isentropic relations:
\begin{eqnarray}
	T_t = 1+\frac{1}{2}(\gamma-1)M_{\infty}^{2} \, & \mbox{and} & 
	P_t = \frac{1}{\gamma}(T_t)^{\frac{\gamma}{\gamma-1}} \, \mbox{.}
	\label{eq:Tot-entry}
\end{eqnarray}
The dimensionless static temperature and pressure are deduced from Eq.\ \eqref{eq:Tot-entry},
resulting in
\begin{eqnarray}
	\left( T \right)_{1,j,k}=\frac{T_t}{1+\frac{1}{2}(\gamma-1)(u^2+v^2+w^2)_{1,j,k}} \, 
	& \mbox{and} & 
	\left( p \right)_{1,j,k}=\frac{1}{\gamma}(T)_{1,j,k}^{\frac{\gamma}{\gamma-1}} \, \mbox{.}
	\label{eq:Stat-entry}
\end{eqnarray}
For the supersonic case, all conserved variables receive jet property values.

The far field boundary conditions are implemented outside of the jet area in order to correctly
propagate information comming from the inner domain of the flow to the outter region of 
the simulation. However, in the present case, $\xi$, instead of $\eta$, as presented in 
the previous subsection, is the normal direction used to define the Riemann invariants.

\subsection{Exit Boundary Condition}

At the exit plane, the same reasoning of the jet entrance boundary is applied. This time, 
for a subsonic exit, the pressure is obtained from the outside and all other variables are 
extrapolated from the interior of the computational domain by a zero-order extrapolation. The 
conserved variables are obtained as
\begin{eqnarray}
	(\rho)_{I_{MAX},j,k} &=& \frac{(p)_{I_{MAX},j,k}}{(\gamma-1)(e)_{I_{MAX}-1,j,k}} \mbox{,} \\
	(\vec{u})_{I_{MAX},j,k} &=& (\vec{u})_{I_{MAX}-1,j,k}\mbox{,} \\
	(e_i)_{I_{MAX},j,k} &=& 
	(\rho)_{I_{MAX},j,k}\left[ (e)_{I_{MAX}-1,j,k}+
	\frac{1}{2}(\vec{u})_{I_{MAX},j,k}\cdot(\vec{u})_{I_{MAX},j,k} \right] \, \mbox{,}
	\label{eq:exit}
\end{eqnarray}
in which $I_{MAX}$ stands for the last point of the mesh in the axial direction. For 
the supersonic exit, all properties are extrapolated from the interior domain.

\subsection{Centerline Boundary Condition}

The centerline boundary is a singularity of the coordinate transformation, and, hence, 
an adequate treatment of this boundary must be provided. The conserved properties 
are extrapolated from the ajacent longitudinal plane and are averaged in the azimuthal 
direction in order to define the updated properties at the centerline of the jet.

\begin{figure}[ht]
       \begin{center}
       {\includegraphics[width=0.5\textwidth]{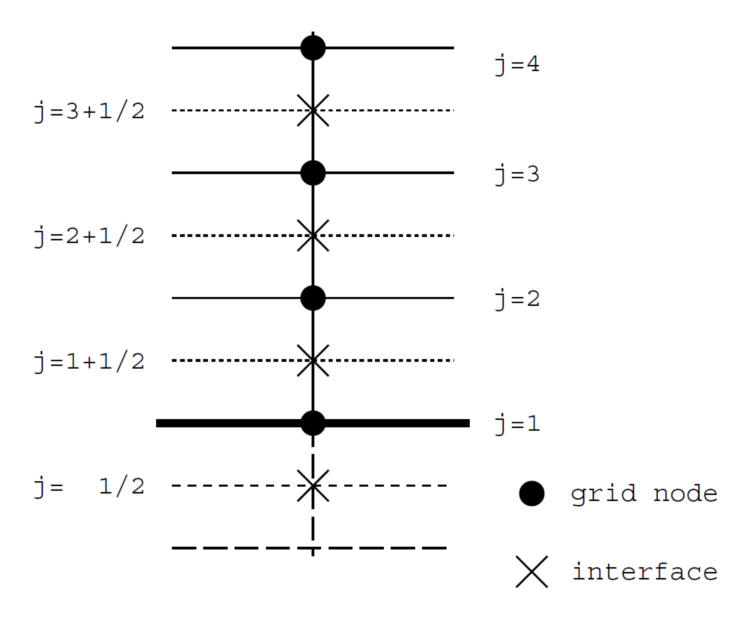}}\\
	   \caption{Boundary points dissipation \cite{BIGA02}.}\label{fig:centerline}
       \end{center}
\end{figure}

The fourth-difference terms of the artificial dissipation scheme, used in the present 
work, are carefully treated in order to avoid the five-point difference stencils at 
the centerline singularity. 
If one considers the flux balance at one grid point near the centerline boundary in 
a certain coordinate direction, let $w_{j}$ denote a component of the $\mathcal{W}$ 
vector from Eq.\ \eqref{eq:W_dissip} and $\mathbf{d}_{j}$ denote the corresponding artificial
dissipation term at the mesh point $j$. In the present example, 
$\left(\Delta w\right)_{j+\frac{1}{2}}$ stands for the difference between the solution
at the interface for the points $j+1$ and $j$. The fouth-difference of the dissipative
fluxes from Eq.\ \eqref{eq:dissip_term} can be written as
\begin{equation}
	\mathbf{d}_{j+\frac{1}{2}} = \left( \Delta w \right)_{j+\frac{3}{2}} 
	- 2 \left( \Delta w \right)_{j+\frac{1}{2}}
	+ \left( \Delta w \right)_{j-\frac{1}{2}} \, \mbox{.}
\end{equation}
Considering the centerline and the point $j=1$, as presented in Fig.\ 
\ref{fig:centerline}, the calculation of $\mathbf{d}_{1+\frac{1}{2}}$ demands the 
$\left( \Delta w \right)_{\frac{1}{2}}$ term, which is unknown since it is outside the
computation domain. In the present work a extrapolation is performed and given by
\begin{equation}
	\left( \Delta w \right)_{\frac{1}{2}} =
	- \left( \Delta w \right)_{1+\frac{1}{2}} \, \mbox{.}
\end{equation}
This extrapolation modifies the calculation of $\mathbf{d}_{1+\frac{1}{2}}$ that can be written as
\begin{equation}
	\mathbf{d}_{j+\frac{1}{2}} = \left( \Delta w \right)_{j+\frac{3}{2}} 
	- 3 \left( \Delta w \right)_{j+\frac{1}{2}} \, \mbox{.}
\end{equation}
The approach is plausible since the centerline region is smooth and does not have high
gradient of properties.

\subsection{Periodic Boundary Condition}

A periodic condition is implemented between the first ($K=1$) and the last point in the 
azimutal direction ($K=K_{MAX}$) in order to close the 3-D computational domain. There 
are no boundaries in this direction, since all the points are inside the domain. The first
and the last points, in the azimuthal direction, are superposed in order to facilitate
the boundary condition implementation which is given by
\begin{eqnarray}
	(\rho)_{i,j,K_{MAX}} &=& (\rho)_{i,j,1} \, \mbox{,} \nonumber\\
	(u)_{i,j,K_{MAX}} &=& (u)_{i,j,1} \, \mbox{,} \nonumber\\
	(v)_{i,j,K_{MAX}} &=& (v)_{i,j,1} \, \mbox{,} \\
	(w)_{i,j,K_{MAX}} &=& (w)_{i,j,1} \, \mbox{,} \nonumber\\
	(e)_{i,j,K_{MAX}} &=& (e)_{i,j,1} \, \mbox{.} \nonumber
	\label{eq:periodicity}
\end{eqnarray}


\section{Mesh Generation}

\begin{figure}[htb!]
	\begin{center}
        \subfigure[Frontal view of the computational domain for 
		the jet flow simulation]{
		\includegraphics[width=0.7\textwidth]
		{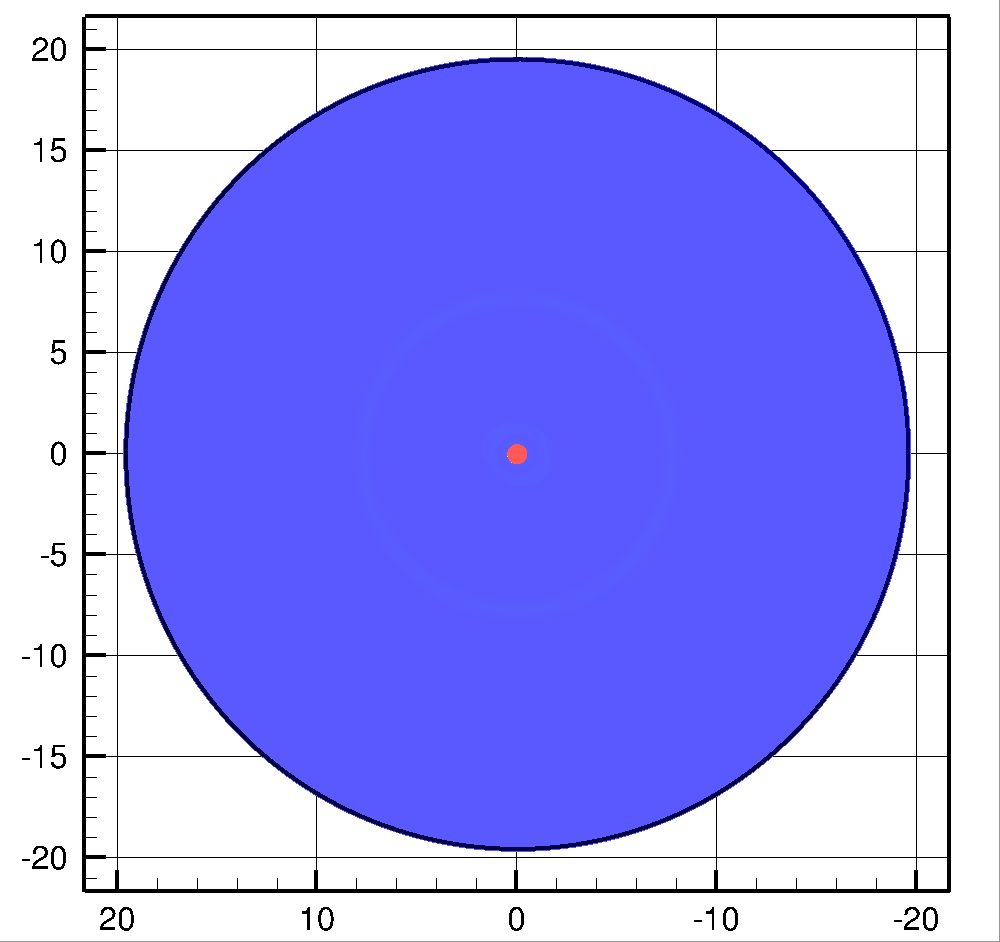}\label{fig:domain}
		}
		\subfigure[2-D view of the mesh generation.]{
		    \includegraphics[width=0.95\textwidth]
			{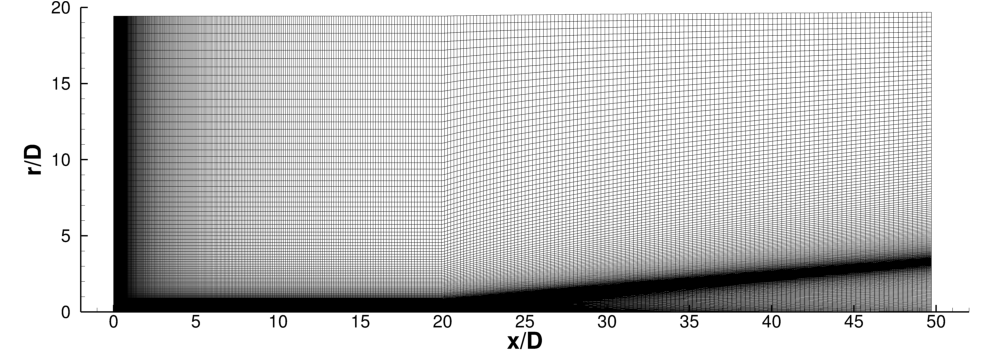}
			\label{fig:2d_mesh}
		}
		\caption{Computational domain and mesh for a jet flow
		simulation} \label{fig:domain_mesh}
	\end{center}
\end{figure}

A structured mesh generator is created in order to provide 
CGNS grid files for the simulations performed in the present 
work. Figure \ref{fig:domain} illustrates a frontal view of a representative 
domain for the jet flow simulations. The blue boundary is the freestream
region and the red one is the entrance of the jet flow. 
The complete 3-D mesh is created by rotating a 2-D mesh around the horizontal 
direction, $x$. A view of this 2-D mesh is illustrated in 
Fig.~\ref{fig:2d_mesh}. This approach generates a singularity at the centerline 
of the domain. The treatment of this region is discussed in the boundary conditions 
section. The authors chose not to include the nozzle geometry and the jet entrance 
is located at $x=0$, between $|r|/d \leq 0.5$, where $|r|$ is the distance from the 
centerline in the radial direction and $d$ is the incoming jet diameter. 
In Fig.~\ref{fig:2d_mesh}, two distinct regions are visible: the developing region 
for $x/d < 20$ and the fully-developed region where the flow is self-similar. In the 
latter, the shear layer has a constant growth rate given by $\mathcal{S} \approx 0.094$ 
\cite{Pope00}. This trend is typical of jet flows and the mesh refinement follows this 
evolution. The refinement is made using hyperbolic tangent functions, with a finest 
grid resolution near the jet entrance and along the slip line of the jet. The mesh 
is coarsened in the far field in order to diffuse the acoustic waves and avoid 
reflections at the farfield domain boundaries.



%% file: sources/hpc/hpc.tex
\section{High Performance Computing}

\subsection{Code Parallelization and Improvements}

The developement of a 3-D finite-difference numerical code within 
the IAE CFD group was initiated in 1997 \cite{Vieira_azevedo_97,
Vieira_azevedo_98,Vieira_azevedo_98_2}. In the sequence, viscous 
terms, multigrid techniques, implicit residual smoothing and turbulence 
modeling were also added to the solver \cite{BIGA02}. 
As previously discussed, large eddy simulations demand 
very dense grids. Therefore, one can state that LES is only possible 
throught multi-processing computing. The authors have invested significant 
efforts to improve the outdated numerical solver toward a more 
sophisticated computational level. The FORTRAN 77 programming language
was replaced by the FORTRAN 90 standard and dynamic memory allocation is 
included in the new version. These improvements are performed in order to 
simplifly the code structure and to enable the implementation of parallel 
computation routines.

Pre-processing and post-processing procedures are very important steps of 
computer-aided engineering. The original version did not have an 
input/output standard accepted by the vizualisation tools used by 
the scientific community. Hence, the CFD general notation system, 
also know by the CGNS \cite{Piorier98,cgns_overview} acronym, is 
implemented in the present work. The main post-processing and 
visualization tools accept the CGNS standard.

\begin{figure}[htb!]
       \begin{center}
       {\includegraphics[width=0.45\textwidth]
	   {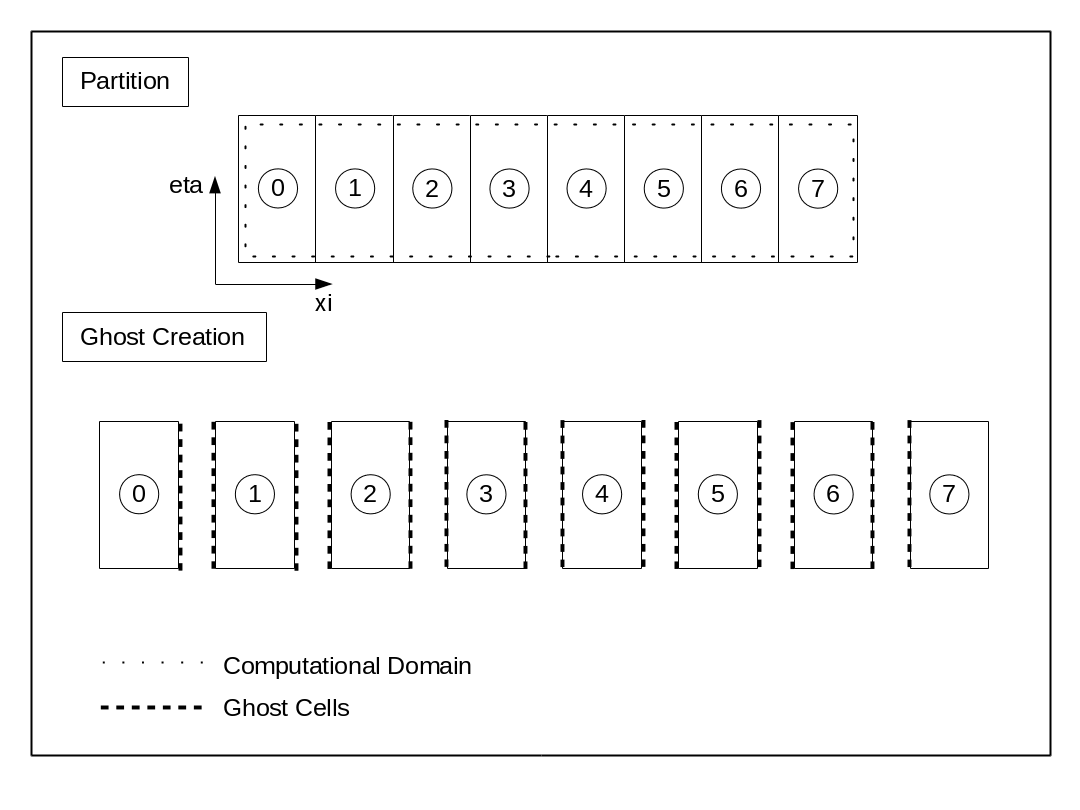}}\\
       \caption{One dimensional mesh partitioning procedure in the axial 
	   direction.}\label{fig:partition}
       \end{center}
\end{figure}

The mesh partitioning is performed when the simulation begins. One 
dimensional and two dimensional partitionings are performed in the present 
work. The former partitionates the domain in the axial direction or the
azimuthal direction and the latter partitionates the domain in both, 
the axial and azymuthal directions. Figure \ref{fig:partition} illustrates 
the 1-D partitioning procedure in the axial direction. Each processor reads 
the complete domain and performs the segmentation. The partition numeration 
starts at zero to be consistent with the message passing interface notation 
which applies the C\# programming language vector standard. This direction, 
in the general curvilinear coordinate system adopted in the present work, 
is the $\xi$ direction for axial partitionig and $\zeta$ for the azimuthal
partitioning. 

Numerical data exchange between the partitions is necessary regarding 
parallel computation. In the present work, the ghost concept is applied. 
Ghost points are added to the local partition at the main flow 
direction and azimuthal direction in order to carry information of the 
neighboring points. The artificial dissipation scheme implemented in the 
code \cite{jameson_mavriplis_86} uses a five points stencil which demands 
information of the two neighbors of a given mesh point. Hence, a two
ghost-point layer is created at the beginning and at the end of each 
partition. 

\begin{figure}[ht]
       \begin{center}
		   \subfigure[Forward communication between 1-D partitions.]{
           \includegraphics[width=0.45\textwidth]
		   {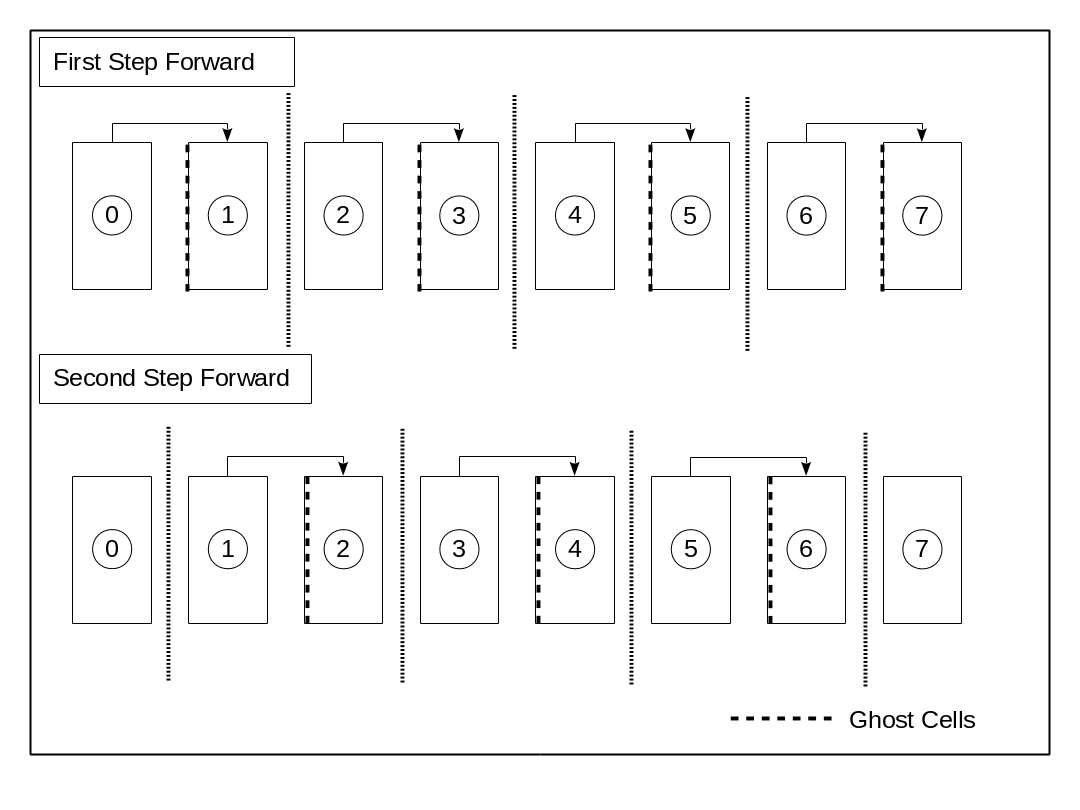} \label{fig:forward}
		   }
		   \subfigure[Backward communication between 1-D partitions.]{
           \includegraphics[width=0.45\textwidth]
		   {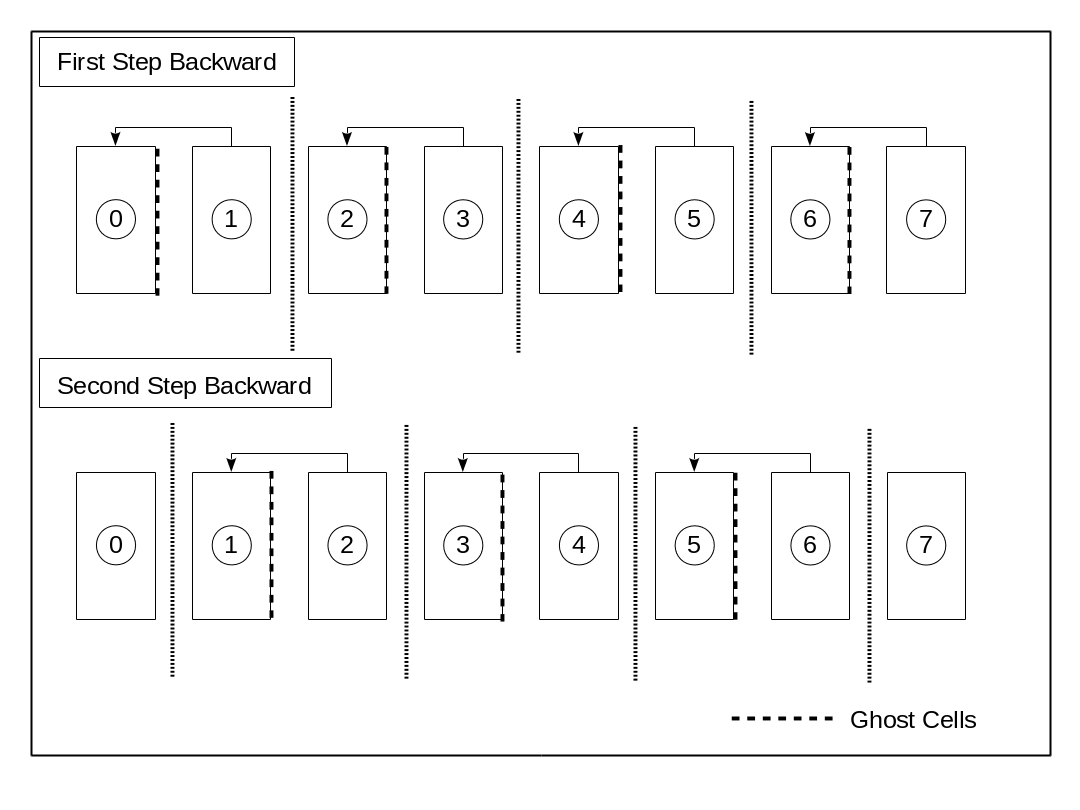} \label{fig:backward}
		   }
		   \caption{1-D communication between partitions.}\label{fig:comm}
       \end{center}
\end{figure}

After the partitioning and the ghost cell creation, each processor performs
the computation. Communication between pairs of partition neighbors are 
performed in order to allow data information pass through the computational 
domain. For each direction, the data exchange is performed in four blocking 
steps. Figures \ref{fig:forward} and \ref{fig:backward} demonstrate the 1-D
communication which initially is performed in the forward direction. Even 
partitions send information of their two last local layers to the ghost 
points at the left of odd partitions. If the last partition is even, 
it does not share information in this step. In the sequence, odd partitions 
send information of their two last local layers to the ghost points at 
the left of even partitions. If the last partition is odd, it does not 
share information in this step. The third and the fourth steps are backward 
communications. First, odd partitions send data of their two first 
local layers to the ghost points at the right of even partitions. Finally, 
all even partitions, but the first one, send data of their two first 
local layers to the ghost points at the right of odd partitions.

\begin{figure}[ht]
       \begin{center}
		   \subfigure[2-D partitioning in the axial and azimuthal direction.]{
           \includegraphics[width=0.45\textwidth]
		   {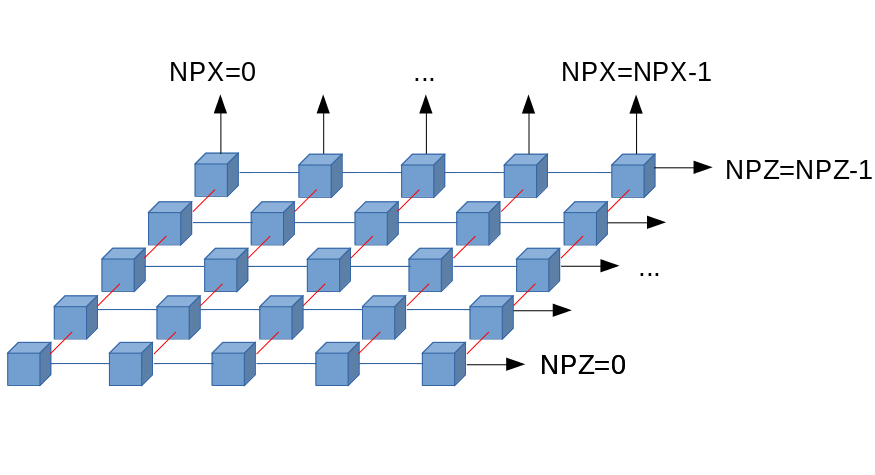} \label{fig:XZ_1}
		   }
		   \subfigure[Mapping of the 2-D communication.]{
           \includegraphics[width=0.45\textwidth]
		   {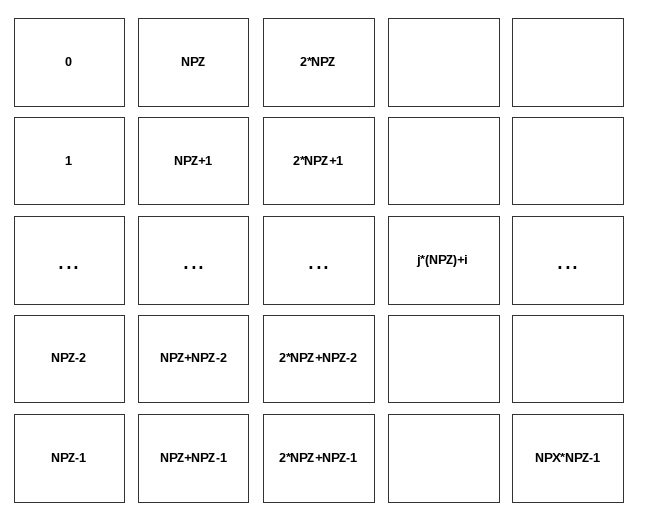} \label{fig:XZ_2}
		   }
		   \caption{2-D partitioning and mapping.}\label{fig:comm_2d}
       \end{center}
\end{figure}

The extention to the 2-D communication is straightforward. Figures \ref{fig:XZ_1} 
and \ref{fig:XZ_2} illutrate the 2-D segmentation and mapping of the domain. 
The numeration of each partition is performed using a matrix index system in which
the rows represent the position in the axial direction and the columns represent
the position in the azimuthal direction. NPX and NPZ denote the number of partitions
in the axial and azimuthal directions, respectively. The 2-D communication is performed 
first in the azimuthal direction and, then, in the axial direction in the same fashion as the 
1-D data exchange is performed.

The user can choose the frequency in which the solution is written in a CGNS file.
After a chosen number of iterations, the main processor collects the data from 
all other processors and appends the solution to the CGNS file in a time-dependent 
fashion. Therefore, after the simulation, it is possible to recover the solution
as a function of the physical time. However, despite the advantages, the output
approach, in which one processor collects data of all other processors, is a serial
output and, consequently, it can reduce the parallel computational performance. 
CGNS provides parallel I/O software which needs to be implemented in the future.


\subsection{Computational Performance Assessment}

Parallel computations can largely decrease the time of CFD simulations. However,
high performance computing upgrade of a serial solver must be carefully performed. 
The parallel version of a solver ought to produce exactly the same results as the 
single processed version. Moreover, the communication between partitions is not 
free and can affect the computational performance in parallel. The partitioning 
of the domain increases the amount of communication between processors which can 
race with the time spent on computation and, consequently, deteriorate the 
performance of the solver.

The speedup is one of the most common figures-of-merit for the performance evaluation of 
parallel algorithms and architectures \cite{Ertel94} and it is used in the present 
work in order to measure the computational efficiency of the parallel solver and 
compare with the ideal case. Different approaches are used by the scientific community
in order to calculate the speedup \cite{gustafson88,xian10}. In the present work,
the speedup, ${Sp}_{N}$, is calculated as %
\begin{equation}
	{Sp}_{N} = \frac{T_{N}}{T_{s}} \, \mbox{,}
	\label{eq:speedup}
\end{equation}
in which $T_{N}$ and $T_{s}$ stand for the time spent to perform one thousand 
iterations using $N$ processors and one single processor, respectively. The 
efficiency as a function of the number of processors, $\eta_{N}$, is written 
considering Amdahl's law \cite{amdahl67} as
%
%
%
\begin{equation}
	\eta_{N} = \frac{{Sp}_{N}}{N} \, \mbox{.}
\end{equation}

Gustafson \cite{gustafson88} has discussed that the efficiency of a parallel task scales 
with the size of the problem. Therefore, different meshes are created in order to evaluate
the performance of the solver as a function of the number of points within partitions. 
Table \ref{tab:mesh} presents the mesh size and the number of points in the three directions 
of the domain, $n_{\xi}$, $n_{\eta}$ and $n_{\zeta}$.
\begin{table}[htbp]
\begin{center}
\caption{Mesh information}
\label{tab:mesh}
\begin{tabular}{|ccccc||ccccc|}
\hline
Mesh & $n_{\xi}$ & $n_{\eta}$ & $n_{\zeta}$ & Mesh size &
Mesh & $n_{\xi}$ & $n_{\eta}$ & $n_{\zeta}$ & Mesh size \\
\hline
1 & 200 & 150 & 40 & 1.20E06 &
6 & 400 & 150 & 10 & 6.00E05 \\
2 & 400 & 150 & 40 & 2.4E06 &
7 & 400 & 150 & 60 & 3.60E06 \\
3 & 800 & 150 & 40 & 4.80E06 &
8 & 800 & 300 & 120 & 2.88E07 \\
4 & 400 & 75 & 40 & 1.20E06 &
9 & 200 & 75 & 40 & 6.00E05 \\
5 & 400 & 300 & 40 & 4.80E06 &
10 & 200 & 75 & 20 & 3.00E05 \\ 
\hline
\end{tabular}
\end{center}
\end{table}
The ratio of computation-communication costs per partition is measured in the present work 
regarding the evaluation of the best partitioning arrangement for the solver. Only the
1-D partitioning in the axial direction is evaluated in the present work. The ratio is 
proportional to the number of points in the main flow direction, $n_{\xi}$. This statement 
is true because the computational cost of a partition is directly related to the size of 
the domain and the communication cost is directly related to the number of ghost-points of 
each partition which, for the present 1-D axial partitioning, is proportional to the number of 
points in the radial direction times the number of points in the azimuthal direction. 
%
%

\begin{figure}[htb]
       \begin{center}
		   \subfigure[Speedup as a function of the 
		   number of processors.]{
		   \includegraphics[width=0.45\textwidth]
		   {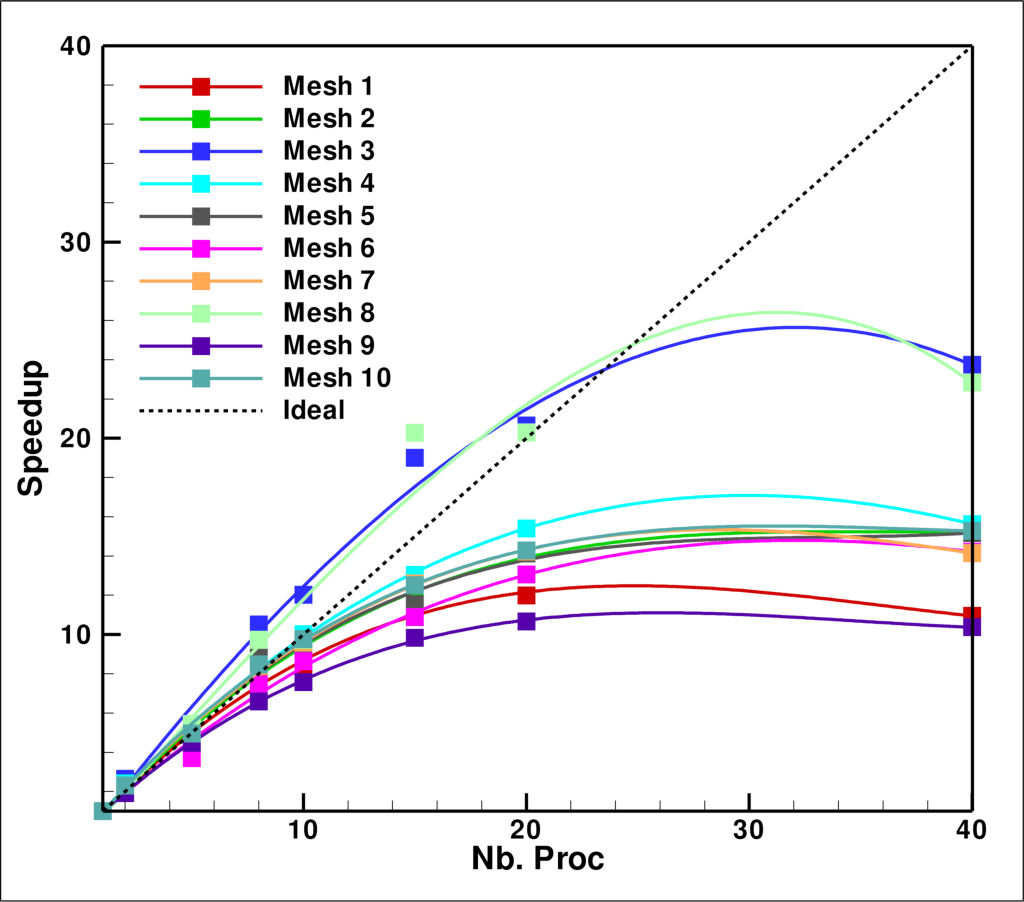}\label{fig:speedup}
		   }
		   \subfigure[Efficiency as a 
		   function of the number of processors.]{
		   \includegraphics[width=0.45\textwidth]
		   {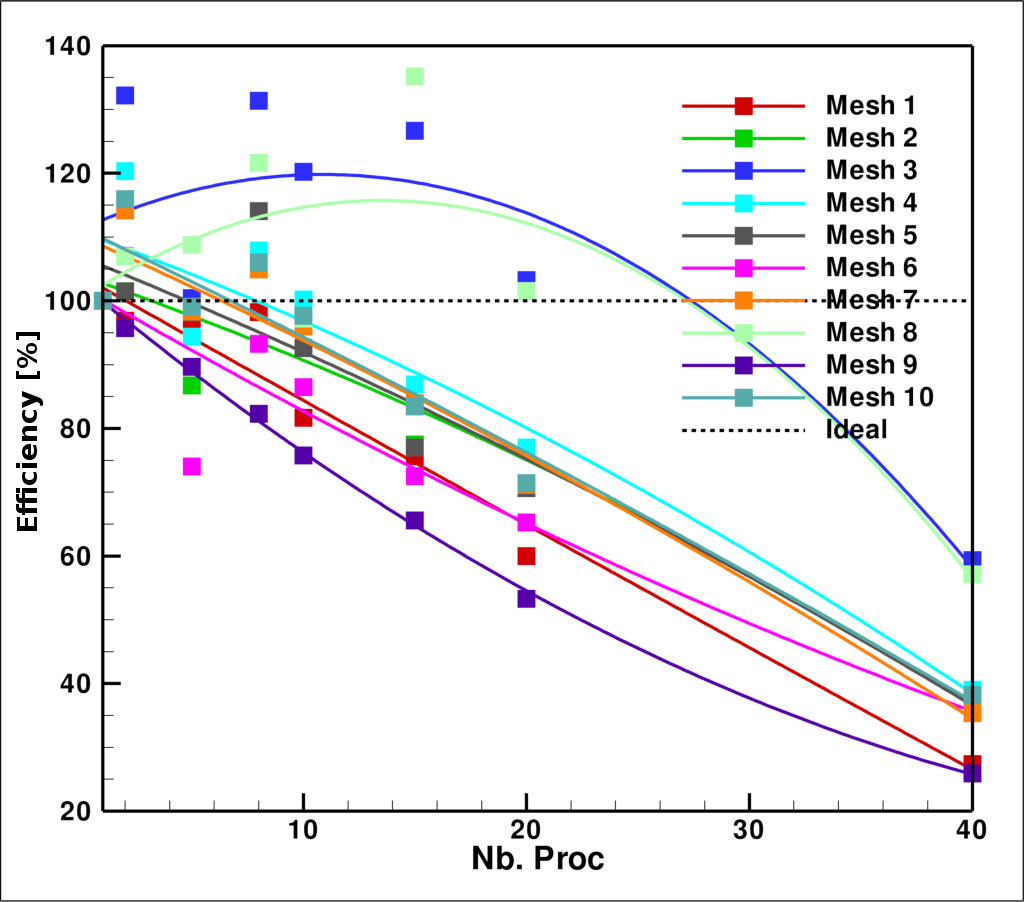}\label{fig:performance}
		   }
		   \caption{Trends observed for speedup and computational efficiency of
		   the parallel version of the code.}\label{fig:speedup_perfm}
       \end{center}
\end{figure}

%
A simplified flow simulation was performed using the meshes presented in Tab.\ 
\ref{tab:mesh} and different number of processors. The speedup and efficiency of 
the solver is presented in Tab.\ \ref{tab:hpc}. In this table, $N$ represents the 
number of processors used in each test case. Figures \ref{fig:speedup} and 
\ref{fig:performance} illustrate the trends for the evolution of speedup and efficiency as 
a function of the number of processors. In these figures, the symbols indicate the actual 
speedup and efficiency values obtained in the various tests, whereas the curves represent 
the trends given by least-squares best fits through the data. 
One can state that the code has the best 
computational performace for $n_{\xi}>50$. The efficiency is $\mathcal{O}(1)$ for all 
configurations with $n_{\xi}>50$ presented in this section. For $40 < n_{\xi} < 50$, 
the communication costs become significant and begin to deteriorate the performance, 
which decreases even further with smaller partition sizes. For $n_{\xi} \approx 5$ the communication 
is more expensive than computation. The speedup coefficient is higher for configurations 
with $n_{\xi} \approx 10$ than configurations with $n_{\xi} \approx 5$. The results  
motivated the extention to a 2-D partitioning in the code. However, the scalability for 
such partition configurations was not measured for the the current paper.
\begin{table}[htbp]
\begin{center}
\caption{Table of speedup and efficiency as a function of the number of processors}
\label{tab:hpc}
\begin{tabular}{|c|c|}
\hline
{\bf Mesh Nb.} & \\ 
\hline
{\bf 1} & 
 \begin{tabular}{c|c|c|c|c|c|c|c}
	$N$ & {\bf 2} & {\bf 5} & {\bf 8} & {\bf 10} & {\bf 15} & {\bf 20} & {\bf 40} \\ 
    \hline
	$S_{N}$ & 1.937 & 4.768 & 7.858 & 8.165 & 11.35 & 12.00 & 10.95 \\
    \hline
	$\eta_{N}$ & 0.969 & 0.953 & 0.982 & 0.816 & 0.756 & 0.600 & 0.274 \\
    \hline
	$n_{\xi}$ & 100 & 40 & 25 & 20 & 13 & 10 & 5
 \end{tabular}\\
\hline
{\bf 2} & 
 \begin{tabular}{c|c|c|c|c|c|c|c}
	$N$ & {\bf 2} & {\bf 5} & {\bf 8} & {\bf 10} & {\bf 15} & {\bf 20} & {\bf 40} \\ 
    \hline
	$S_{N}$ & 2.030 & 4.338 & 9.125 & 9.265 & 11.62 & 14.15 & 15.23 \\
    \hline
	$\eta_{N}$ & 1.015 & 0.868 & 1.141 & 0.926 & 0.775 & 0.708 & 0.381 \\
    \hline
	$n_{\xi}$ & 200 & 80 & 50 & 40 & 26 & 20 & 10
 \end{tabular}\\
\hline
{\bf 3} & 
 \begin{tabular}{c|c|c|c|c|c|c|c}
	$N$ & {\bf 2} & {\bf 5} & {\bf 8} & {\bf 10} & {\bf 15} & {\bf 20} & {\bf 40} \\ 
    \hline
	$S_{N}$ & 2.644 & 5.021 & 10.51 & 12.02 & 19.00 & 20.64 & 23.75 \\
    \hline
	$\eta_{N}$ & 1.322 & 1.004 & 1.314 & 1.212 & 1.266 & 1.032 & 0.594 \\
    \hline
	$n_{\xi}$ & 400 & 160 & 100 & 80 & 53 & 40 & 20
 \end{tabular}\\
\hline
{\bf 4} & 
 \begin{tabular}{c|c|c|c|c|c|c|c}
	$N$ & {\bf 2} & {\bf 5} & {\bf 8} & {\bf 10} & {\bf 15} & {\bf 20} & {\bf 40} \\ 
    \hline
	$S_{N}$ & 2.406 & 4.718 & 8.632 & 10.02 & 13.04 & 15.40 & 15.62 \\
    \hline
	$\eta_{N}$ & 1.203 & 0.944 & 1.079 & 1.002 & 0.869 & 0.770 & 0.390 \\
    \hline
	$n_{\xi}$ & 200 & 80 & 50 & 40 & 26 & 20 & 10
 \end{tabular}\\
\hline
{\bf 5} & 
 \begin{tabular}{c|c|c|c|c|c|c|c}
	$N$ & {\bf 2} & {\bf 5} & {\bf 8} & {\bf 10} & {\bf 15} & {\bf 20} & {\bf 40} \\ 
    \hline
	$S_{N}$ & 2.030 & 4.909 & 9.124 & 9.257 & 11.55 & 14.13 & 15.17 \\
    \hline
	$\eta_{N}$ & 1.015 & 0.982 & 1.140 & 0.926 & 0.770 & 0.706 & 0.379 \\
    \hline
	$n_{\xi}$ & 200 & 80 & 50 & 40 & 26 & 20 & 10
 \end{tabular}\\
\hline
{\bf 6} & 
 \begin{tabular}{c|c|c|c|c|c|c|c}
	$N$ & {\bf 2} & {\bf 5} & {\bf 8} & {\bf 10} & {\bf 15} & {\bf 20} & {\bf 40} \\ 
    \hline
	$S_{N}$ & 2.141 & 3.703 & 7.461 & 8.648 & 10.88 & 13.06 & 14.22 \\
    \hline
	$\eta_{N}$ & 1.070 & 0.741 & 0.933 & 0.865 & 0.725 & 0.653 & 0.355 \\
    \hline
	$n_{\xi}$ & 200 & 80 & 50 & 40 & 26 & 20 & 10
 \end{tabular}\\
\hline
{\bf 7} & 
 \begin{tabular}{c|c|c|c|c|c|c|c}
	$N$ & {\bf 2} & {\bf 5} & {\bf 8} & {\bf 10} & {\bf 15} & {\bf 20} & {\bf 40} \\ 
    \hline
	$S_{N}$ & 2.300 & 4.916 & 8.393 & 9.599 & 12.60 & 14.24 & 14.15 \\
    \hline
	$\eta_{N}$ & 1.150 & 0.983 & 1.050 & 0.960 & 0.840 & 0.712 & 0.354 \\
    \hline
	$n_{\xi}$ & 200 & 80 & 50 & 40 & 26 & 20 & 10
 \end{tabular}\\
\hline
{\bf 8} & 
 \begin{tabular}{c|c|c|c|c|c|c|c}
	$N$ & {\bf 2} & {\bf 5} & {\bf 8} & {\bf 10} & {\bf 15} & {\bf 20} & {\bf 40} \\ 
    \hline
	$S_{N}$ & 2.140 & 5.441 & 9.730 & 9.762 & 20.27 & 20.30 & 22.84 \\
    \hline
	$\eta_{N}$ & 1.070 & 1.088 & 1.216 & 0.973 & 1.352 & 1.015 & 0.571 \\
    \hline
	$n_{\xi}$ & 400 & 160 & 100 & 80 & 53 & 40 & 20
 \end{tabular}\\
\hline
{\bf 9} & 
 \begin{tabular}{c|c|c|c|c|c|c|c}
	$N$ & {\bf 2} & {\bf 5} & {\bf 8} & {\bf 10} & {\bf 15} & {\bf 20} & {\bf 40} \\ 
    \hline
	$S_{N}$ & 1.915 & 4.482 & 6.585 & 7.576 & 9.838 & 10.66 & 10.37 \\
    \hline
	$\eta_{N}$ & 0.957 & 0.896 & 0.823 & 0.758 & 0.656 & 0.530 & 0.259 \\
    \hline
	$n_{\xi}$ & 100 & 40 & 25 & 20 & 13 & 10 & 5
 \end{tabular}\\
\hline
{\bf 10} & 
 \begin{tabular}{c|c|c|c|c|c|c|c}
	$N$ & {\bf 2} & {\bf 5} & {\bf 8} & {\bf 10} & {\bf 15} & {\bf 20} & {\bf 40} \\ 
    \hline
	$S_{N}$ & 2.319 & 4.953 & 8.480 & 9.768 & 12.52 & 14.29 & 15.29 \\
    \hline
	$\eta_{N}$ & 1.116 & 0.990 & 1.060 & 0.977 & 0.835 & 0.714 & 0.382 \\
    \hline
	$n_{\xi}$ & 100 & 40 & 25 & 20 & 13 & 10 & 5
 \end{tabular}\\
\hline
\end{tabular}
\end{center}
\end{table}

\begin{figure}[htb]
       \begin{center}
		   {
		   \includegraphics[width=0.6\textwidth]
		   {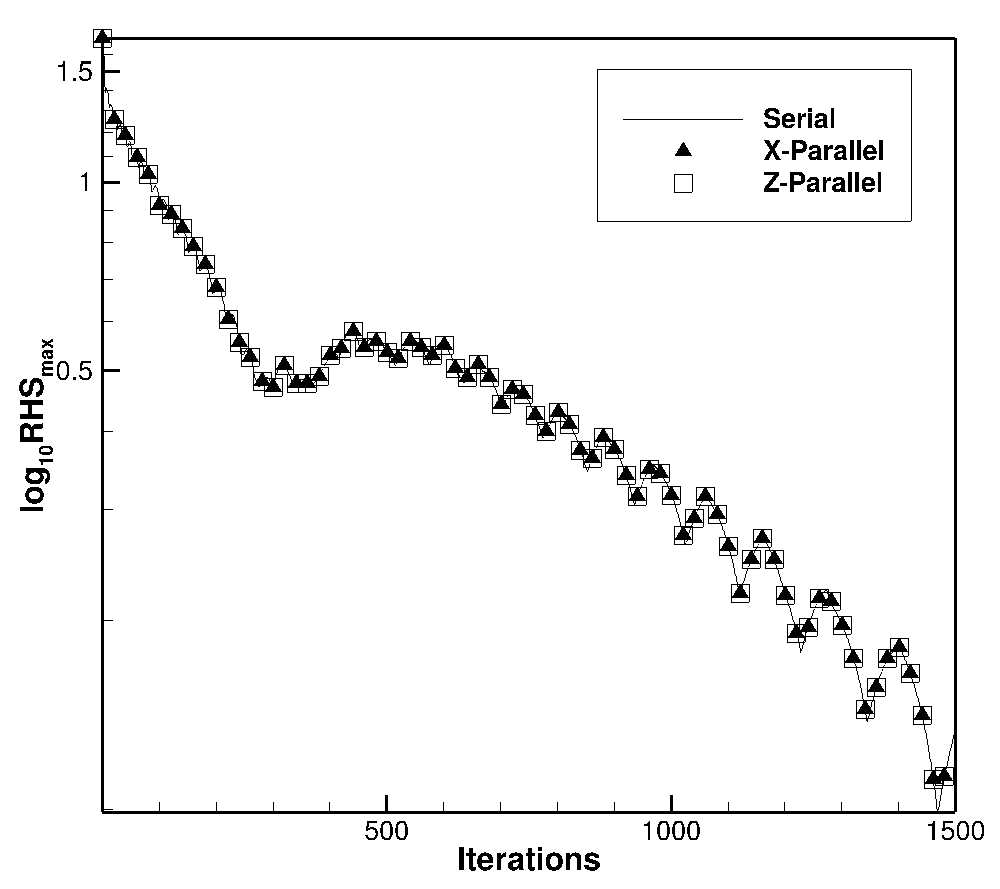}
		   }
		   \caption{Maximum RHS of large eddy simulations of jet flows
		   using different partitioning.}\label{fig:repro}
       \end{center}
\end{figure}

The binary reproducibility of the solver was also evaluated in the present work.
A high performance computing code must produce exactly the same results as the serial 
version of the code, regardless of
the number of processors used. Therefore, three large eddy simulations of jet flows
were performed using a single processor, 1-D partitioning in the axial direction 
and 1-D partitioning in the azimuthal direction. Figure \ref{fig:repro} illustrates
the history of maximum absolute value of the right-hand side of the continuity equation 
through 1500 iterations. In other words, the figure is showing what would be the 
$L_{\infty}$ norm of the residue of the continuity equation if these were standard 
steady state calculations. The goal is the evaluation of differences between serial and
parallel configurations. Visually, all the three simulations present the same
behavior, independently of the number of processors. However, only the partitioning 
in the axial direction presents binary reproducibility compared the serial simulation. 
The azimuthal partitioning has accumulated differences on fifth significant figure at the 
end of the 1500 iterations. This is due to the fact that MPI collective communications 
have to be performed in the azimuthal direction in order to average properties at the 
centerline. Floating-point operations, such as sums and products, are commutative and 
not associative. Thus, the order in which they are performed makes a difference to the 
final outcome \cite{balaji2013}, and, for each azimuthal partitioning configuration, the 
result of the average at the centerline is different in the context of bitwise 
reproducibility. 

%% file: sources/results/results.tex
\section{Numerical Results}

The present section is devoted to a preliminary study of a supersonic 
perfectly expanded jet flow. Results are compared with analytical, 
numerical and experimental data from the literature 
\cite{Bodony05i8,Mendez10,Tanna77,bridges2008turbulence,lo2012,laufer1976}. 
First LES results using the Smagorinsky SGS turbulence closure 
\cite{Smagorinsky63} are discussed. The flow is characterized by an 
unheated perfectly expanded inlet jet with a Mach number of 
$1.4$ at the domain entrance. Therefore, the pressure ratio, 
$PR = P_{j}/P_\infty$, and the temperature ratio, $TR=T_{j}/T_\infty$, 
between the jet exit and the ambient freestream are equal to one, 
$PR = 1$ and $TR=1$. The time step used in the simulation is constant 
and equal to $2.5 \times 10^{-5}$ in dimensionless form. The Reynolds number 
of the jet is $Re = 1.57 \times 10^{6}$. The radial and longitudinal 
dimensions of the smallest cell of the computational domain are given by 
$(\Delta \underline{r})_{min}=0.002$ and $(\Delta \underline{x})_{min}=0.0126$, 
respectively, again in dimensionless form. This cell is located in the shear 
layer of the jet, at the entrance of the computational domain. The number 
of points in the azimuthal direction is $N_\theta=180$. The mesh domain is 
composed by 14.4 million points. The present mesh reproduces well the 
main geometric characteristics of the reference grid\cite{Mendez10}.

\begin{figure}[htb!]
	\begin{center}
		\subfigure[]{
		\includegraphics[width=0.475\textwidth]
		{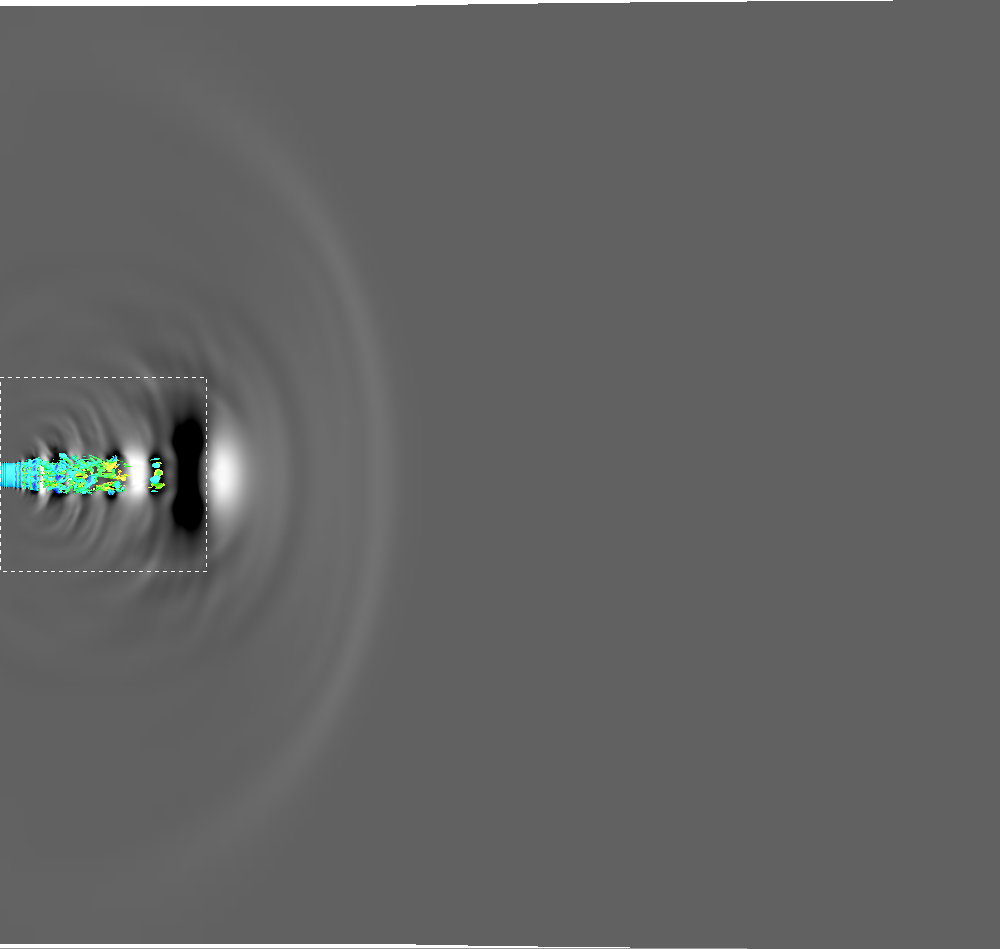}\label{fig:vort-p}
		}
		\subfigure[]{
		\includegraphics[width=0.475\textwidth]
		{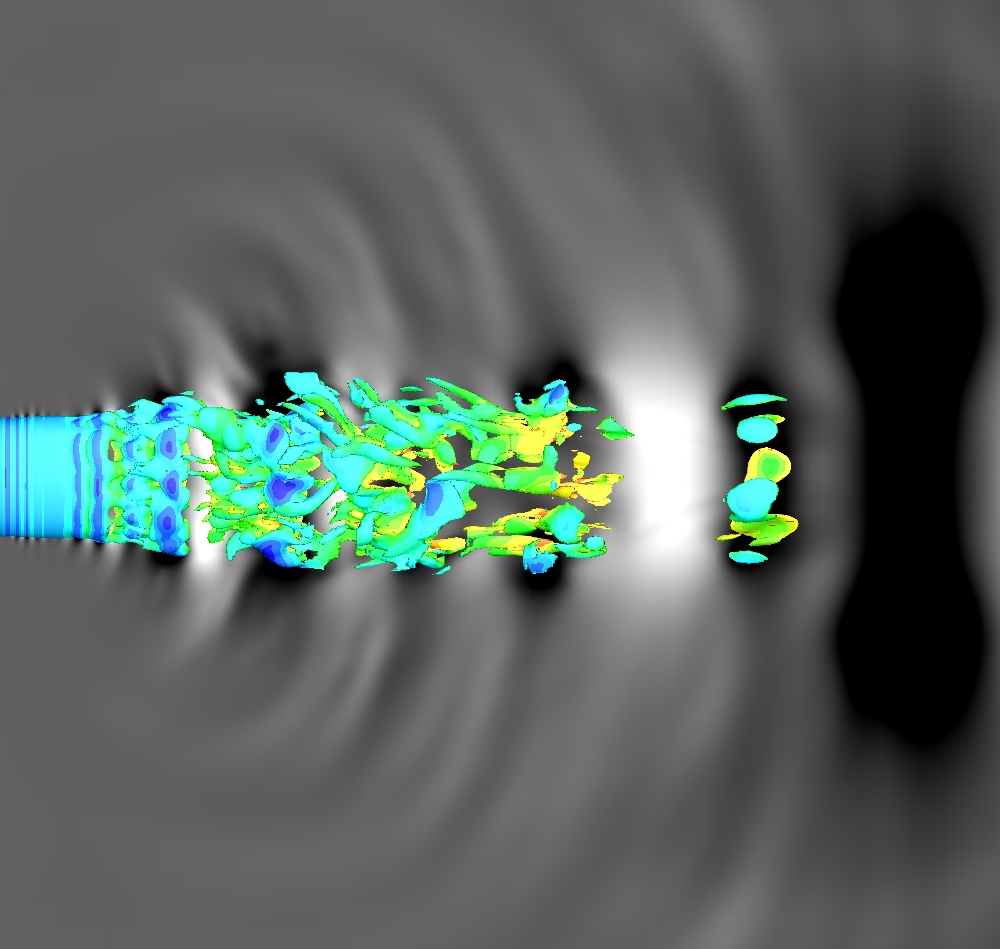}\label{fig:vort-p-zoom}
		}
	\end{center}
    \caption{Instantaneous pressure and vorticity fields for the perfectly expanded supersonic jet.
	The gray scale contours show $\underline{p}$ from 0.4 to 0.9, while a dimensionless iso-contour of vorticity 
	of magnitude $|\omega|D/U_{j}=10$ is colored by the streamwise velocity $\underline{u}$ from -0.2 (blue) to 1.4 (red).}
    \label{fig:vort-p-fields}
\end{figure}

Generally, compressible jet flow studies performed in the literature
present mean field properties. However, as the present simulation is 
not statistically converged yet, mean quantities could not be calculated. 
As an example, using the same flow configuration, Mendez\cite{Mendez10} 
needed 300 characteristic times, $\mathcal{T}_\infty$, to obtain averages. 
Here, $\mathcal{T}_\infty$ is defined as the ratio between the diameter D and the 
ambient sound speed $a_\infty$. The simulation time of the cited reference corresponds approximatively to 
four flow through times. One flow through time is the time for a particle 
to cross the entire domain from the jet entrance to the domain exit.
The reader should remember that, as $TR=1$ and $a_j = a_\infty$, the  
characteristic time in the present work is equal to that reported in the work of Mendez 
\cite{Mendez10}. In the present work, the code assessment of the supersonic jet 
flow is made by observing the last obtained snapshot. This snapshot corresponds 
to 18.5 $\mathcal{T}_j$ which is far less than the computational time in the reference. However, the 
authors chose to start the analysis of the jet flow using this field. Figure 
\ref{fig:vort-p} displays instantaneous fields of pressure and vorticity of 
the jet at that instant. The jet is entering into the domain at the left 
and evolves towards the right of the domain. As the flow has been initialized from
stagnation conditions and not, for instance, from a converged RANS solution, a front pressure
shock wave is preceding the first vortical structures. Figure \ref{fig:vort-p-zoom} 
focuses on the jet flow entrance. At the beginning of the mixing layer, thin 
axisymmetrical vortices are growing and pairing, leading then to turbulent 
transition at approximatively $x=1.5D$. This figure also displays Mach waves 
emitted by the mixing layer, displaying a downstream directivity, retrieving 
traditional trends \cite{Bodony05i8,Mendez10,Tanna77,bridges2008turbulence,lo2012,laufer1976}. 
One can notice that the strongest acoustic sources 
match the coherent structure locations, identifiable by the condensed vorticity 
rings.

\begin{figure}[htb!]
	\begin{center}
		\subfigure[]{
		\includegraphics[width=0.475\textwidth]
		{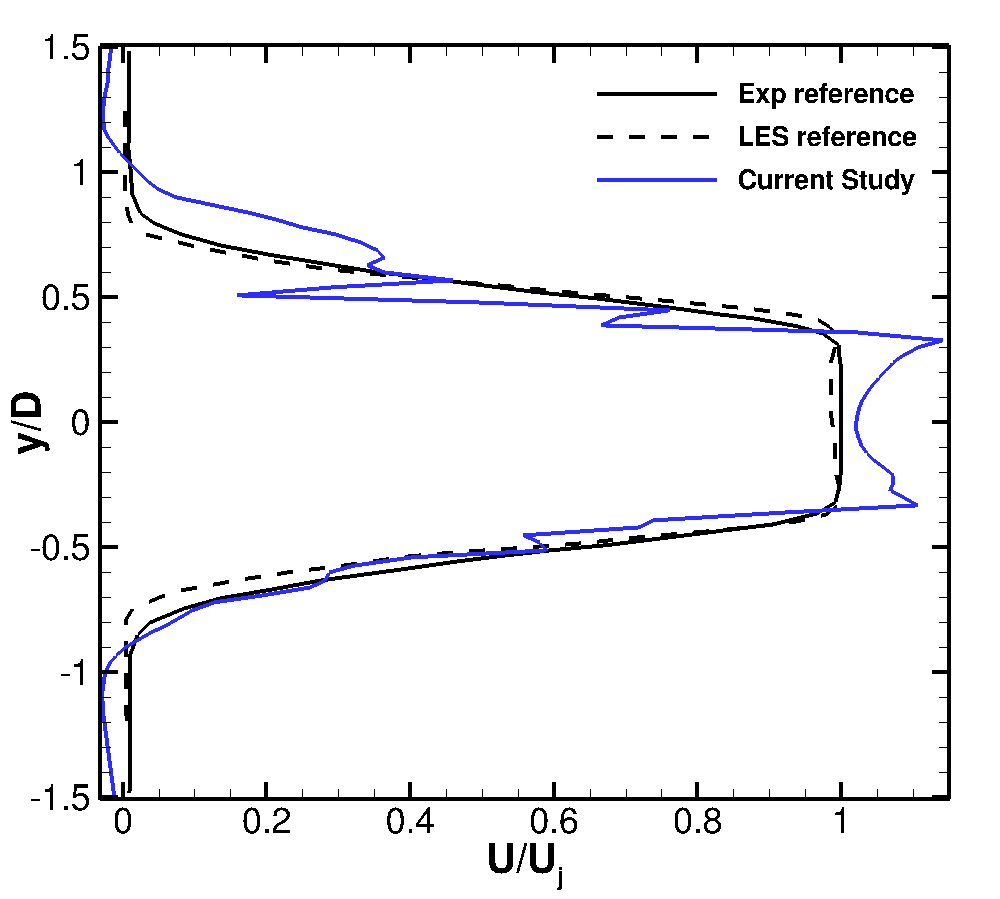}\label{fig:prof-comp}
		}
		\subfigure[]{
		\includegraphics[width=0.475\textwidth]
		{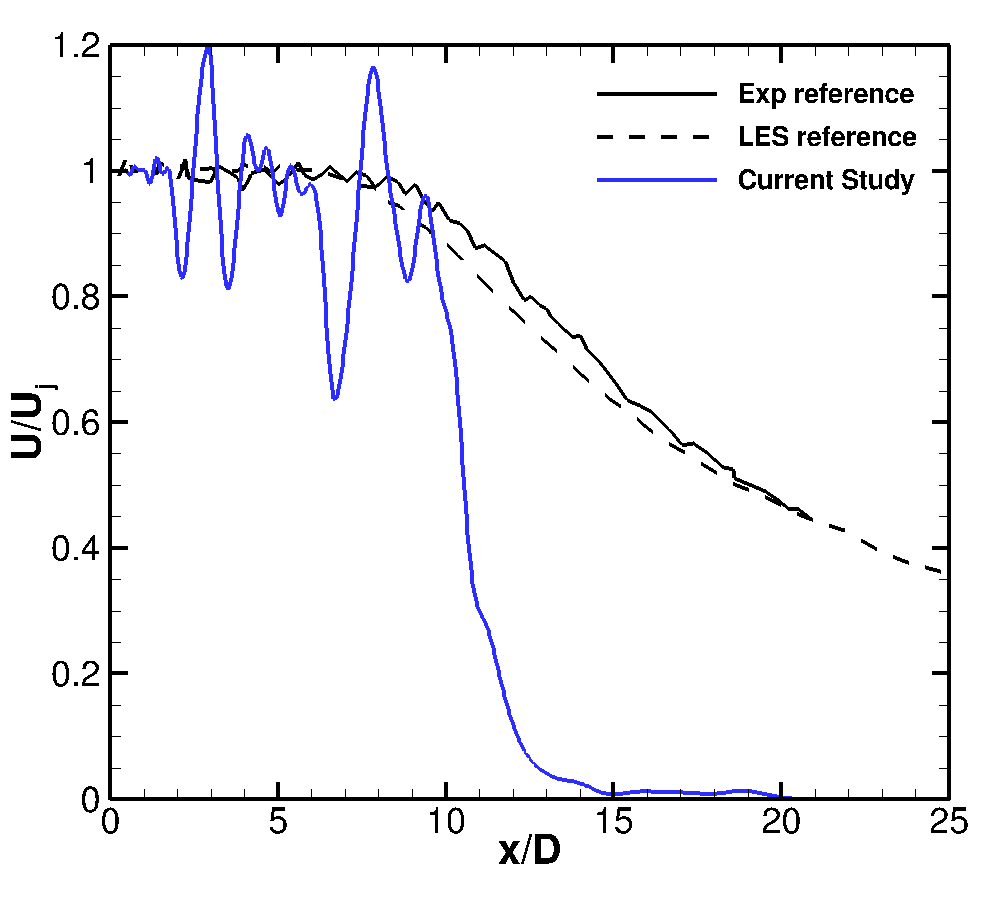}\label{fig:stream-comp}
		}
	\end{center}
    \caption{Comparison between numerical and experimental streamwise velocity component profiles along
	the centerline ($a$), and radial profiles at $x=2.5D$. Instantaneous results from the present study
	are compared with LES results\cite{Mendez10} and experimental 	ones\cite{bridges2008turbulence}.}
    \label{fig:vel-comp}
\end{figure}

Figure \ref{fig:prof-comp} presents the radial profile of the streamwise 
velocity component at $x=2.5D$ and compares it with the  
averaged numerical and experimental profiles from Mendez\cite{Mendez10} and 
Bridges\cite{bridges2008turbulence}, respectively.
The centerline values of the velocity profile 
are overestimated in the current simulation when 
compared to the references. This difference is expected since a flat inlet boundary profile is used for
this simulation while a laminar profile, necessarily thicker, is used in the numerical reference work.
However, keeping in mind that the current profile is instantaneous, a fair agreement is obtained.
Similar observations can be made on the results comparing the centerline streamwise velocity components,
shown in Figure \ref{fig:stream-comp}.
The values in the potential core, corresponding to the jet area until $U=0.95 \cdot U_j$,
compare well with the references.
Moreover, one can expect that, as the jet will grow downstream, the 
decreasing slope after the potential core will raise to meet the reference curves, confirming thus
the good trend of the present simulation.

\begin{figure}[htb!]
	\begin{center}
		\includegraphics[width=0.75\textwidth]
		{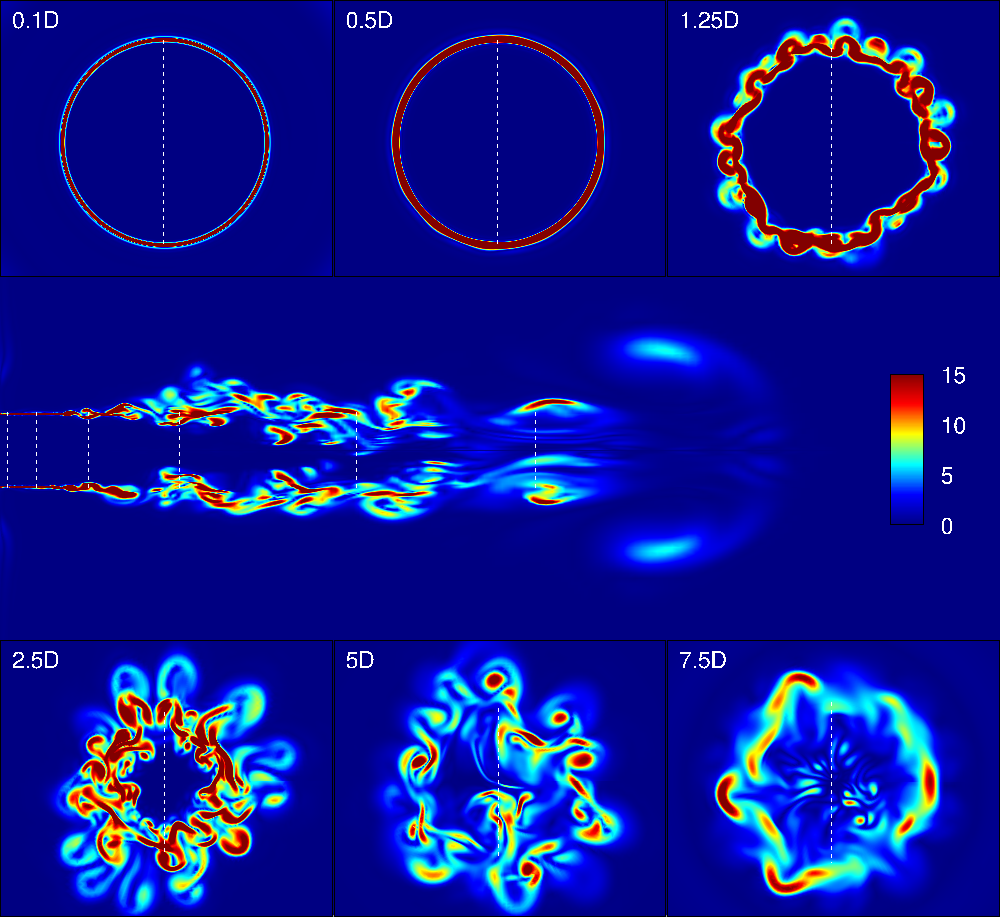}
	\end{center}
	\caption{Instantaneous vorticity magnitude for the present simulation at $M=1.4$.
	}
	\label{fig:jet-slices}
\end{figure}

Another visualization of this jet flow is shown in Figure \ref{fig:jet-slices}. It is possible to observe the 
evolution of the vorticity looking at different cross-flow planes from $\underline{x}=0.1$ to $7.5$. 
At the moment of this snapshot,
slices located after the last streamwise position shown are irrelevant as the jet is not fully developed yet.
From $\underline{x}=0.1$ to 0.25, the mixing layer is getting thicker and it is axisymmetric as already 
observed in Figure \ref{fig:vort-p-fields}.
Further downstream, the flow becomes completely three dimensional and is no longer axisymmetric, transitioning
towards a turbulent state. Still, the turbulent vortical structures are larger than already observed in the
literature\cite{Mendez10,bres2012towards}. This is due to the low order 
of the computational code and the barely acceptable refinement of the grid for LES simulations.
The mixing layer continues widening as one moves downstream. One can clearly expect that the mixing 
layer will spread 
until the jet centerline near the end of the potential core.

\begin{figure}[htb!]
	\begin{center}
		\subfigure[$\underline{u}$]{
		\includegraphics[width=0.475\textwidth]
		{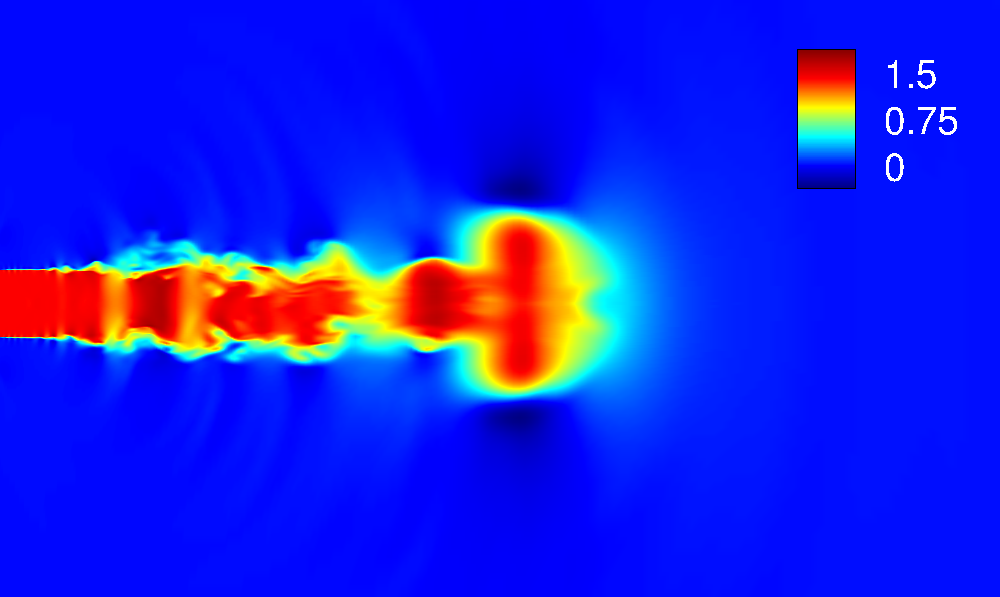}\label{fig:u}
		}
		\subfigure[$\underline{\rho}$]{
		\includegraphics[width=0.475\textwidth]
		{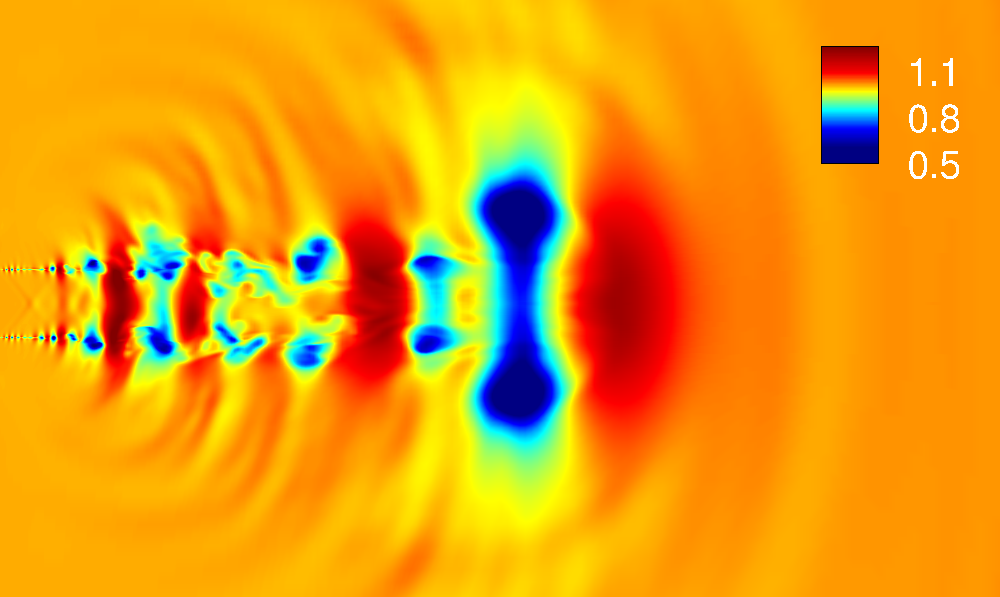}\label{fig:rho}
		}
		\subfigure[$\underline{v}$]{
		\includegraphics[width=0.475\textwidth]
		{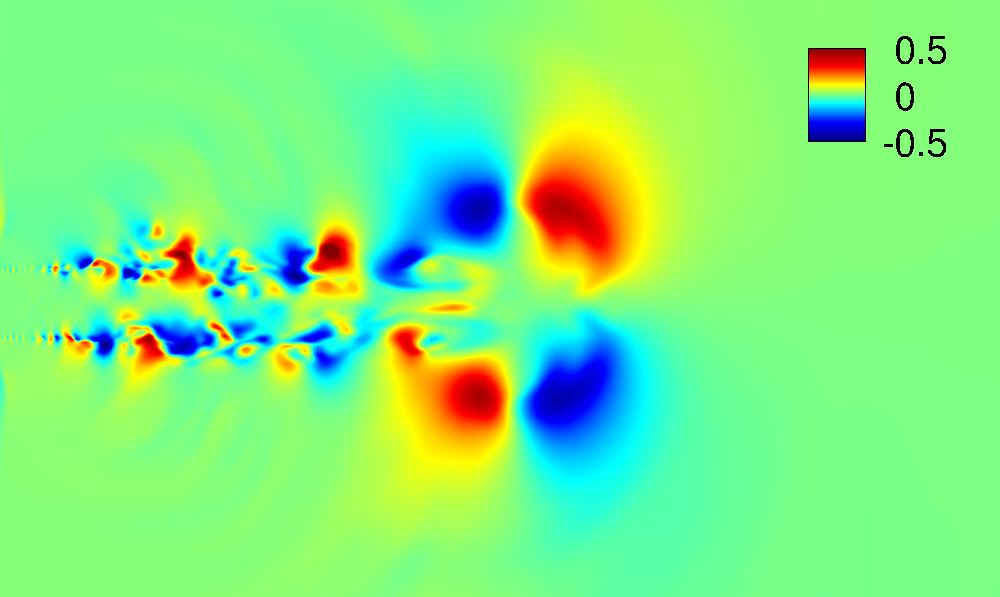}\label{fig:v}
		}
		\subfigure[$\underline{E}$]{
		\includegraphics[width=0.475\textwidth]
		{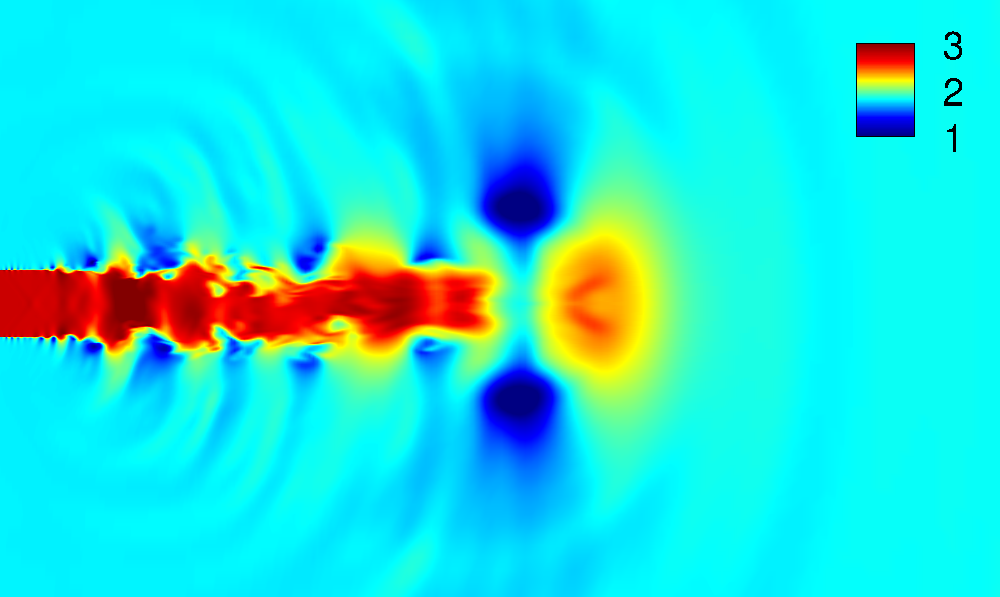}\label{fig:E}
		}
		\subfigure[$\underline{w}$]{
		\includegraphics[width=0.475\textwidth]
		{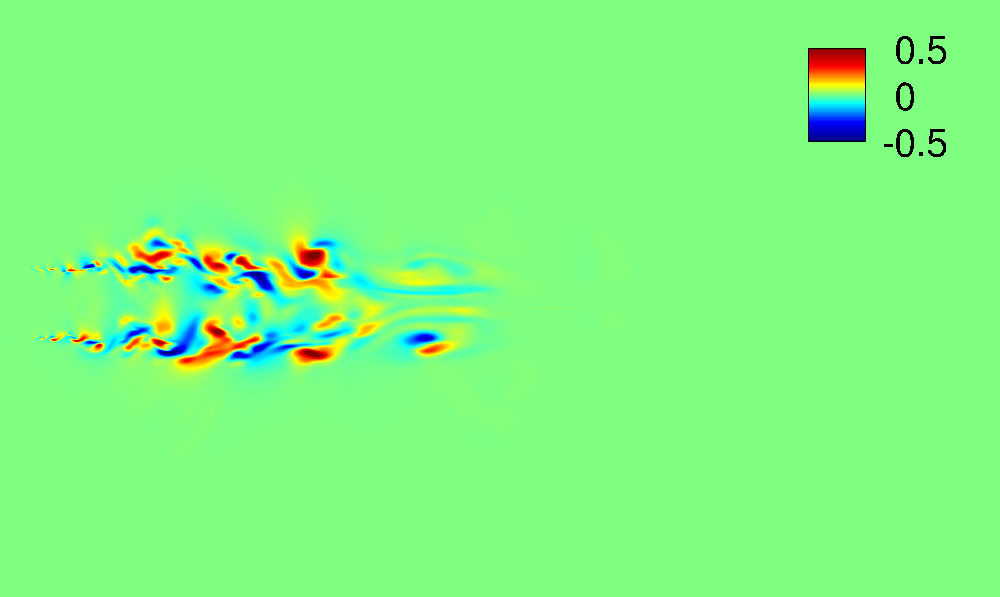}\label{fig:w}
		}
		\subfigure[$\underline{p}$]{
		\includegraphics[width=0.475\textwidth]
		{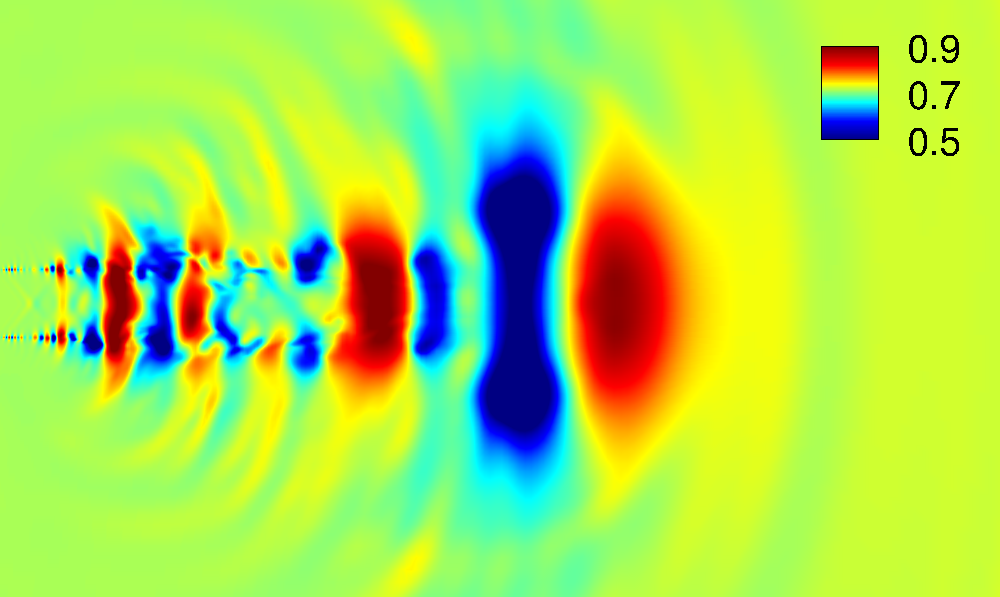}\label{fig:p}
		}
	\end{center}
	\caption{Instantaneous properties of the perfectly expanded 
	supersonic jet flow.}
	\label{fig:jet-prop}
\end{figure}

Finally, flow properties are displayed in Figure \ref{fig:jet-prop}.
In this two dimensional snapshot, radial and azimuthal velocity components correspond to vertical 
and spanwise velocity components, respectively. The $\underline{v}$ and $\underline{w}$ 
velocity components
are of the same order of magnitude in the jet mixing layer turbulent region, illustrating thus the 
three dimensional turbulent behaviour of the jet mixing layer.
Looking at the second column of Figure \ref{fig:jet-prop}, one can observe weak shock cells in 
the instantaneous density, total energy and pressure fields. Downstream directivity patterns are
very similar to those seen in the reference work \cite{Mendez10, bres2012towards}, confirming 
one more time that the current simulation is evolving towards the correct trends.

%% file: sources/conclusion/conclusion.tex
\section{Concluding Remarks}

The current work is based on the development of novel 
large eddy simulation tool for the study of supersonic
jet flows. The numerical code background is a RANS 
solver designed for rocket flow configurations. The 
original solver is improved in order to simulate 
supersonic jet flows using an LES formulation within a
multiprocessor environment.

%

Parallel and serial simulations are performed in order to validate 
the reproducibility of the code. Two different approaches are used 
to partitionate the computational domain for the parallel simulation. 
One partitioning is performed in the axial direction and the other one 
in the azimuthal direction. All simulations have produced similar results. 
However, binarywise reproducibility was only achieved for the axial 
partitioning. There are differences on the last significant figure for the 
azimuthal partitioning. The centerline condition demands the use of 
collective MPI communication when the azimuthal partitioning is performed.
Such communications can destroy the binarywise reproducibility of the solver. 
Efforts are being taken in order to solve the issue without impacting
the performance of the parallel solver. The 1-D computational efficiency 
in the axial direction of the parallel code, using different numbers of 
processors, is also assessed and discussed. The code has presented linear 
speedups for partition width with more than 50 points in the flow 
direction. This behavior is expected since the communication between 
partitions starts to race against computation for smaller domains.
However, due to the large size of the problem, the 2-D partitioning strategy
has been used for all LES calculations here. 

Large eddy simulations of a perfectly expanded supersonic jet 
flow configuration are performed. The classical Smagorinsky
model is used to calculate the subgrid terms. 
Results obtained in this work do not allow the calculation of statistically time-averaged properties 
since the simulation only ran during a very limited time period. However, since the aim of the paper is 
mostly to validate the LES computational tool, the authors believe that the results here included 
should provide enough information to allow evaluation of the trends of jet physics. 
Early quantitative comparisons with numerical \cite{Mendez10} and
experimental \cite{bridges2008turbulence} reference work are obtained. 
The jet mixing layer is showing strong instabilities evolving towards a turbulent state.
The classical downstream directivity pattern already observed in the 
literature \cite{Bodony05i8,Mendez10,Tanna77,bridges2008turbulence,laufer1976} is retrieved.
The velocity profiles also show a good agreement with the references.

Finally, regarding the low order, more dissipative numerical scheme, the early results 
show a fairly good agreement with the references, indicating that grid refinement could 
probably resolve such issues. 
Hence, for future simulations, a much more refined new grid should be used 
to counterbalance the dissipation induced by the numerical scheme.
In addition, the computational code needs to be further improved through the availability 
of more sophisticated time-marching schemes.
For instance, the time step used in the present simulations is one hundred times smaller 
than the one used in literature \cite{Mendez10}, leading to a prohibitive simulation cost. 
Hence, a new time integration method should be implemented, such as a preconditioned dual-time step
procedure \cite{pandya2003implementation}.
Furthermore, acoustic predictions of the far field noise using the Ffowcs Williams and Hawkings
analogy will be performed using the reference work of Wolf {\it et al.}\cite{Wolf11}. Results will be
compared one more time with the numerical studies of Mendez {\it et al.}\cite{Mendez10},
Br\`{e}s {\it et al.}\cite{bres2012towards} and Khalighi {\it et al.}\cite{khalighi2010unstructured}.
The influence of the inlet boundary profile on the acoustics will also be investigated.

%% file: main.bbl
\begin{thebibliography}{10}
\newcommand{\enquote}[1]{``#1''}

\bibitem{Bodony05i8}
Bodony, D. and Lele, S.~K., \enquote{On Using Large-Eddy Simulation For the
  Prediction of Noise From Cold and Heated Turbulent Jets,} {\em Physics of
  Fluids\/}, Vol.~17, No.~8, Aug. 2005.

\bibitem{Garnier09}
Garnier, E., Adams, N., and Sagaut, P., {\em Large Eddy Simulation for
  Compressible Flows\/}, Springer, 2009.

\bibitem{Wolf2012}
Wolf, W.~R., Azevedo, J. L.~F., and Lele, S.~K., \enquote{Convective Effects
  and the Role of Quadrupole Sources for Aerofoil Aeroacoustics,} {\em Journal
  of Fluid Mechanics\/}, Vol.~708, 2012, pp.~502--538.

\bibitem{BIGA02}
Bigarella, E. D.~V., {\em Three-Dimensional Turbulent Flow Over Aerospace
  Configurations\/}, {M.Sc.} {T}hesis, Instituto Tecnol\'{o}gico de
  Aeron\'{a}utica, S\~ao Jos\'e dos Campos, SP, Brasil, 2002.

\bibitem{Vreman1995}
Vreman, A.~W., {\em Direct and Large-Eddy Simulation of the Comperssible
  Turbulent Mixing Layer\/}, Ph.D. thesis, Universiteit Twente, 1995.

\bibitem{Smagorinsky63}
Smagorinsky, J., \enquote{General Circulation Experiments with the Primitive
  Equations: I. The Basic Experiment,} {\em Monthly Weather Review\/}, Vol.~91,
  No.~3, March 1963, pp.~99--164.

\bibitem{eidson85}
Eidson, T.~M., \enquote{Numerical Simulation of the Turbulent
  Rayleigh--B{\'e}nard Problem Using Subgrid Modelling,} {\em Journal of Fluid
  Mechanics\/}, Vol.~158, 1985, pp.~245--268.

\bibitem{Mendez10}
Mendez, S., Shoeybi, M., Sharma, A., Ham, F.~E., Lele, S.~K., and Moin, P.,
  \enquote{Large-Eddy Simulations of Perfectly-Expanded Supersonic Jets:
  Quality Assessment and Validation,} {\em {\em AIAA Paper No.\ 2010--0271}\/},
  January 2010.

\bibitem{Tanna77}
Tanna, H.~K., \enquote{An Experimental Study of Jet Noise Part {I}: Turbulent
  Mixing Noise,} {\em Journal of Sound and Vibration\/}, Vol.~50, No.~3, 1977,
  pp.~405--428.

\bibitem{bridges2008turbulence}
Bridges, J. and Wernet, M.~P., \enquote{Turbulence Associated with Broadband
  Shock Noise in Hot Jets,} {\em AIAA paper\/}, Vol.~2834, 2008, pp.~2008.

\bibitem{Boussinesq1877}
Boussinesq, M.~J., \enquote{Essai sur la Th\'{e}orie des Eaux Courantes,} {\em
  In: M\'{e}moires Pr\'{e}sent\'{e}s par Divers Savants \`{a} l'Academie des
  Sciences, tome {XXIII}\/}, Imprimerie Nationale, 1877, pp. 43--47.

\bibitem{Erlebacher92}
Erlebacher, G., Hussaini, M.~Y., Speziale, C.~G., and Zang, T.~A.,
  \enquote{Toward the Large-Eddy Simulation of Compressible Turbulent Flows,}
  {\em Journal of Fluid Mechanics\/}, Vol.~238, 1992, pp.~155--185.

\bibitem{Lilly65}
Lilly, D.~K., \enquote{On the Computational Stability of Numerical Solutions of
  Time- Dependent Non-Linear Geophysical Fluid Dynamics Problems,} {\em Monthly
  Weather Review\/}, Vol.~93, No.~1, January 1965, pp.~11--25.

\bibitem{Lilly67}
Lilly, D.~K., \enquote{The Representation of Small-Scale Turbulence in
  Numerical Simulation Experiments,} {\em {\em IBM Form No. 320-1951},
  Proceedings of the IBM Scientific Computing Symposium on Environmental
  Sciences\/}, Yorktown Heights, N.Y., 1967, pp. 195--210.

\bibitem{moin91}
Moin, P., Squires, K., Cabot, W., and Lee, S., \enquote{A Dynamic Subgrid-Scale
  Model for Compressible Turbulence and Scalar Transport,} {\em Physics of
  Fluids A: Fluid Dynamics (1989-1993)\/}, Vol.~3, No.~11, 1991,
  pp.~2746--2757.

\bibitem{Martin00}
Mart\'{i}n, M.~P., Piomelli, U., and Candler, G.~V., \enquote{Subgrid-Scale
  Models for Compressible Large-Eddy Simulations,} {\em Theoretical and
  Computational Fluid Dynamics\/}, Vol.~13, No.~5, 2000, pp.~361--376.

\bibitem{lenormand2000}
Lenormand, E., Sagaut, P., Phuoc, L.~T., and Comte, P., \enquote{Subgrid-Scale
  Models for Large-Eddy Simulations of Compressible Wall Bounded Flows,} {\em
  AIAA Journal\/}, Vol.~38, No.~8, 2000, pp.~1340--1350.

\bibitem{coleman1995}
Coleman, G.~N., Kim, J., and Moser, R.~D., \enquote{A Numerical Study of
  Turbulent Supersonic Isothermal-Wall Channel Flow,} {\em Journal of Fluid
  Mechanics\/}, Vol.~305, 1995, pp.~159--183.

\bibitem{lesieur08}
Lesieur, M., {\em {Turbulence in Fluids}\/}, Springer, Saint Martin
  d\'{}H\`{e}res, France, 4th ed., 2008.

\bibitem{larcheveque03}
Larch\^{e}veque, M.~L., {\em Simulation des Grandes \'{E}chelles de
  l`\'{E}coulement Au-Dessus d`une Cavit\'{e}\/}, Ph.D. thesis, Universit\'{e}
  Paris VI - Pierre et Marie Curie, Paris, France, December 2003.

\bibitem{Turkel_Vatsa_1994}
Turkel, E. and Vatsa, V.~N., \enquote{{Effect of Artificial Viscosity on
  Three-Dimensional Flow Solutions},} {\em AIAA Journal\/}, Vol.~32, No.~1,
  1994, pp.~39--45.

\bibitem{jameson_mavriplis_86}
Jameson, A. and Mavriplis, D., \enquote{Finite Volume Solution of the
  Two-Dimensional Euler Equations on a Regular Triangular Mesh,} {\em AIAA
  Journal\/}, Vol.~24, No.~4, Apr. 1986, pp.~611--618.

\bibitem{Jameson81}
Jameson, A., Schmidt, W., and Turkel, E., \enquote{Numerical Solutions of the
  Euler Equations by Finite Volume Methods Using Runge-Kutta Time-Stepping
  Schemes,} {\em {\em AIAA Paper 81--1259}, Proceedings of the {AIAA} 14th
  Fluid and Plasma Dynamic Conference\/}, Palo Alto, Californa, USA, June 1981.

\bibitem{Long91}
Long, L.~N., Khan, M., and Sharp, H.~T., \enquote{{A Massively Parallel
  Three-Dimensional Euler/Navier-Stokes Method},} {\em AIAA Journal\/},
  Vol.~29, No.~5, 1991, pp.~657--666.

\bibitem{Pope00}
Pope, S.~B., {\em Turbulent Flows\/}, Cambridge University Press, Cambridge,
  UK, 2000.

\bibitem{Vieira_azevedo_97}
Vieira, R., Azevedo, J. L.~F., and {Fico Jr.}, N. G. C.~R., \enquote{Slotted
  Transonic Wind Tunnel Flow Simulations Using the Euler Equations,} {\em
  Proceedings of the 14th Brazilian Congress of mechanical Engineering -- COBEM
  97\/}, Bauru, SP, Brazil, Dec. 1997.

\bibitem{Vieira_azevedo_98}
Vieira, R., Azevedo, J. L.~F., {Fico Jr.}, N. G. C.~R., and Basso, E.,
  \enquote{Three Dimensional Simulations of the Flow in a Slotted Transonic
  Wind Tunnel,} {\em Proceedings of the 10h International Conference on Finite
  Element Methods\/}, Tucson, AZ, USA, Jan. 1998, pp. 431--436.

\bibitem{Vieira_azevedo_98_2}
Vieira, R., Azevedo, J. L.~F., {Fico Jr.}, N. G. C.~R., and Basso, E.,
  \enquote{Three Dimensional Flow Simulation in the Test Section of a Slotted
  Transonic Wind Tunnel,} {\em {\em ICAS Paper No. 98-R.3.11}, Proceedings of
  the 21th Congress of the International Council of the Aeronautical
  Sciences\/}, Melbourne, Australia, Sept. 1998.

\bibitem{Piorier98}
Piorier, D. and Enomoto, F.~Y., \enquote{The CGNS System,} {\em {\em AIAA Paper
  No.\ 98--3007}\/}, 1998.

\bibitem{cgns_overview}
{CGNS} -- {CFD} Data Standard, {\em CFD General Notation System Overview and
  Entry-Level Documentation\/}, Document Version 3.1.2.

\bibitem{Ertel94}
Ertel, W., \enquote{On the Definition of Speedup,} {\em {PARLE'94} Parallel
  Architectures and Languages Europe\/}, Springer, Berlin, 1994, pp. 289--300.

\bibitem{gustafson88}
Gustafson, J.~L., \enquote{Reevaluating Amdahl's Law,} {\em Communications of
  the ACM\/}, Vol.~31, No.~5, 1988, pp.~532--533.

\bibitem{xian10}
Sun, X.-H. and Chen, Y., \enquote{Reevaluating Amdahl's Law in Multicore Era,}
  {\em J. Parallel Distrib. Comput.\/}, Vol.~70, No.~2, Feb. 2010,
  pp.~183--188.

\bibitem{amdahl67}
Amdahl, G.~M., \enquote{Validity of the Single Processor Approach to Achieving
  Large Scale Computing Capabilities,} {\em {AFIPS} Conference Proceedings\/},
  Vol.~30, ACM, Atlantic City, N.J., USA, Apr. 1967, pp. 483--485.

\bibitem{balaji2013}
Balaji, P. and Kimpe, D., \enquote{On the Reproducibility of MPI Reduction
  Operations,} {\em 2013 IEEE 10th International Conference on High Performance
  Computing and Communications \& IEEE International Conference on Embedded and
  Ubiquitous Computing (HPCC\_EUC)\/}, IEEE, 2013, pp. 407--414.

\bibitem{lo2012}
Lo, S.~C., Aikens, K.~M., Blaisdell, G.~A., and Lyrintzis, A.~S.,
  \enquote{Numerical Investigation of 3-D Supersonic Jet Flows using Large-Eddy
  Simulation,} {\em International Journal of Aeroacoustics\/}, Vol.~11, No.~7,
  2012, pp.~783--812.

\bibitem{laufer1976}
Laufer, J., Schlinker, R., and Kaplan, R., \enquote{Experiments on Supersonic
  Jet Noise,} {\em AIAA Journal\/}, Vol.~14, No.~4, 1976, pp.~489--497.

\bibitem{bres2012towards}
Bres, G.~A., Nichols, J.~W., Lele, S.~K., and Ham, F.~E., \enquote{Towards Best
  Practices for Jet Noise Predictions with Unstructured Large Eddy
  Simulations,} {\em {\em AIAA Paper No.\ 2012-2965}, Proceedings of the 42nd
  AIAA Fluid Dynamics Conference and Exhibit\/}, New Orleans, Louisiana, June
  2012.

\bibitem{pandya2003implementation}
Pandya, S.~A., Venkateswaran, S., and Pulliam, T.~H., \enquote{Implementation
  of Preconditioned Dual-Time Procedures in {OVERFLOW},} {\em {\em AIAA Paper
  No.\ 2003-0072}, Proceedings of the 41st AIAA Aerospace Sciences Meeting \&
  Exhibit\/}, Reno, NV, Jan. 2003.

\bibitem{Wolf11}
Wolf, W.~R. and Lele, S.~K., \enquote{Aeroacoustic Integrals Accelerated by
  Fast Multipole Method,} {\em AIAA Journal\/}, Vol.~49, No.~7, July 2011,
  pp.~1466--1477.

\bibitem{khalighi2010unstructured}
Khalighi, Y., Ham, F., Moin, P., Lele, S.~K., Colonius, T., Schlinker, R.~H.,
  Reba, R.~A., and Simonich, J., \enquote{Unstructured Large Eddy Simulation
  Technology for Prediction and Control of Jet Noise,} {\em ASME Turbo Expo
  2010: Power for Land, Sea, and Air\/}, American Society of Mechanical
  Engineers, 2010, pp. 57--70.

\end{thebibliography}
